\documentclass[superscriptaddress,amsmath,amssymb,aps,prfluids,10pt,showkeys]{revtex4-2}
\usepackage{amssymb}
\usepackage{changepage}
\usepackage[utf8]{inputenc}
\usepackage{amsmath}
\usepackage{xcolor}
\usepackage[normalem]{ulem}
\usepackage{graphicx}
\usepackage{bm}
\usepackage[margin=1in]{geometry}
\usepackage{fancyhdr}

\newcommand{\der}[2]{\frac{\mathrm{d} #1}{\mathrm{d} #2}}
\newcommand{\vt}{v_\mathrm{t}}
\newcommand{\avt}{|v_\mathrm{t}|}
\newcommand{\vm}{u_\mathrm{rms}}
\newcommand{\rc}{r_\mathrm{c}}
\newcommand{\tc}{t_\mathrm{c}}
\newcommand{\St}{St}
\newcommand{\pf}{\mathrm{p_f}}
\newcommand{\pe}{\mathcal{P}_\mathrm{e}}
\begin{document}

\newcommand{\rev}[1]{{\color{black} #1}}
\newcommand\skrt{\bgroup\markoverwith{\textcolor{red}{\rule[0.5ex]{2pt}{0.4pt}}}\ULon}

\pagestyle{fancy}
\fancyhf{}
\lhead{Settling of particles in thermal convection}
\rhead{Submitted to Phys.~Rev.~Fluids}

\title{Settling of inertial particles in turbulent Rayleigh-B\'enard convection}

\author{Vojt\v{e}ch Pato\v{c}ka}
 \email{patocka.vojtech@gmail.com}
 \affiliation{Institute for Planetary Research, German Aerospace Center (DLR), Berlin, Germany}

\author{Enrico Calzavarini}
 \affiliation{Univ. Lille, ULR 7512 - Unité de Mécanique de Lille - Joseph Boussinesq (UML), F-59000 Lille, France}

\author{Nicola Tosi}
 \affiliation{Institute for Planetary Research, German Aerospace Center (DLR), Berlin, Germany}

\begin{abstract}

The settling behaviour of small inertial particles in turbulent convection is a fundamental problem across several disciplines, from geophysics to metallurgy. In a geophysical context, the settling of dense crystals controls the mode of solidification of magma chambers and planetary-scale magma oceans, while rising of light bubbles of volatiles drives volcanic outgassing and the formation of primordial atmospheres. Motivated by these geophysical systems, we perform a systematic numerical study on the settling rate of particles in a rectangular two-dimensional Rayleigh-B\'{e}nard system with Rayleigh number up to $10^{12}$ and Prandtl number from 10 to 50. Under the idealized condition of spherically-shaped particles with small Reynolds number, two limiting behaviours exist for the settling velocity. On the one hand, Stokes' law applies to particles with small but finite response time, leading to a constant settling rate. On the other hand, particles with a vanishing response time are expected to settle at an exponential rate. Based on our simulations, we present a new physical model that bridges the gap between the above limiting behaviours by describing the sedimentation of inertial particles as a random process with two key components: i) the transport of particles from vigorously convecting regions into sluggish, low-velocity “piles” that naturally develop at the horizontal boundaries of the system, and ii) the probability that particles escape such low-velocity regions without settling at their base. In addition, we identify four distinct settling regimes and analyze the horizontal distribution of sedimented particles. For two of these regimes settling is particularly slow and the distribution is strongly non-uniform, with dense particles being deposited preferentially below major clusters of upwellings. Finally, we apply our results to the crystallization of a magma ocean. Our prediction of the characteristic settling times is consistent with fractional crystallization, i.e. with the efficient separation of dense crystals from the residual lighter fluid. In absence of an efficient mechanism to re-entrain settled particles, equilibrium crystallization appears possible only for particles with extremely small density contrasts.
\end{abstract}


\maketitle


\section{Introduction}
\label{sec:Intro}
\noindent

The motion and sedimentation of small particles in a convecting fluid is of great interest to fluid dynamicists, geophysicists, as well as to metallurgists. Dispersion of pollutants, dust \citep[][]{Schwaiger2012}, and organic material such as pollen in the atmosphere \citep[][]{Helbig2004}, phytoplankton population dynamics in oceans and lakes \citep{Ruiz2004,Squires1995}, crystallization of magma chambers \citep{Martin1989,Koyaguchi1990}, of primordial magma oceans in rocky planets \citep[e.g.][]{ElkinsTanton2012,Solomatov2015}, and of planetary metallic cores \citep[e.g.][]{Breuer2015,Jones2015} are just a few examples of geophysical processes that to some extent can be described through the  settling of inertial particles in a fluid undergoing highly vigorous convection. In industrial settings, the purification of alloys \citep{Lifeng2011} and microfluidic heat transfer technologies \citep[e.g.][]{Chang2008} count into this group.

From a general point of view, the phenomenology of particle-laden turbulent flows has been the subject of extensive studies over the last decades. Researches have focused on the statistical characterisation of particles' dispersion and accumulation upon varying the flow turbulence intensity, the particles' inertia (which is linked to their size and mass density), and their shape \citep{Toschi2009,Voth2017}. For what concerns particle turbulent settling, there are nowadays solid numerical and experimental evidences that particle sinking is enhanced by the effect of turbulence, while the opposite happens for particle rising \citep[for a review, see][]{Mathai2019}. However, these results have been obtained in the context of idealized flows: kinematic flows \citep{Maxey1987, Pasquero2003}, unbounded turbulence \citep{Wang1993}, or channel and pipe flow geometries. Much less explored is the context of thermally-driven flows \citep[for experimental studies, see][]{Martin1989, Lavorel2009}.

Despite the relevance of the problem of particles settling in turbulent flows for a variety of natural and industrial systems, this work is largely motivated by the study of the crystallization of primordial magma oceans, with our choice of the parameter space being inspired by this system (Section \ref{sec:modpar}). Planetary-scale volumes of liquid silicates are thought to form during the accretion and differentiation of terrestrial bodies like the Earth, Mercury, Venus, Mars and the Moon \citep[e.g.,][]{Safronov1969,Tonks1992,Tonks1993}. The way magma oceans solidify is of primary importance for the long-term thermo-chemical evolution of the interior of planets \citep[e.g.,][]{Tosi2020}. Whether or not newly formed crystals settle or remain suspended by turbulent flow determines the initial distribution of the composition of the silicate mantle \citep[e.g.,][]{ElkinsTanton2012}. This is a difficult problem that depends on the density contrast between crystals and melt, the melt viscosity, the size of the crystals and the convective dynamics of the system. If dense crystals are efficiently maintained in suspension, a magma ocean undergoes equilibrium (or batch) crystallization, which leads to a largely homogeneous composition of the rocky mantle. By contrast, if crystals tend to settle and crystal-melt separation is efficient, fractional crystallization takes place. Residual melts are progressively more and more enriched in so-called incompatible elements such as iron-oxides and heat-producing elements. This process ultimately leads to a compositionally-stratified mantle whose long-term evolution can be dramatically different from that of a homogeneous one \citep[e.g.,][]{Tosi2013,Plesa2014,Ballmer2017,Maurice2017}. Magma chambers are small-scale analogs of magma oceans. Upon cooling and solidification these typically undergo strong fractionation, which is often attributed to crystal settling \citep[e.g.,][]{Marsh1985,Martin1988,Martin1989,Koyaguchi1990}.

Similar to the settling of negatively buoyant crystals, floating of light particles is also a fundamental process in the context of the crystallization of magma oceans and magma chambers. In fact, it is the basic mechanism underlying magma degassing, where gas bubbles are released from volatile-saturated magma \citep[e.g.,][]{Sparks2003}. Greenhouse volatiles such as H$_2$O and CO$_2$ also behave as incompatible species and tend to be strongly enriched in the liquid phase upon magma crystallization. The efficiency with which these are released from a magma ocean controls the formation of primordial atmospheres and the timescale of magma ocean solidification \citep[e.g.][]{ElkinsTanton2008,Lebrun2013,Nikolaou2019}.

In the context of a crystallizing magma, \citet{Marsh1985} modelled the transport of particles by convective motions as a turbulent diffusion process, which was a common approach in studies of mixing in turbulent flows \citep[e.g.][]{Bartlett1969, Huppert1980}. Fundamental laboratory experiments aimed at assessing settling rates in a cooling magma were later performed by \citet{Martin1989}, who employed the turbulent diffusion theory to explain their measurements. Assuming the concentration of particles to be spatially uniform, they derived a simple model for particles with a vanishing Stokes velocity according to which the number of suspended particles decays exponentially with time. Although \citet{Martin1989} anticipated that for particles with a larger Stokes velocity their ``diffusion model of turbulent transport will begin to break down and other assumptions will no longer be valid, in particular the assumption of one-dimensionality'', surprisingly little effort has been devoted to extend their work. To our knowledge, no experimental or numerical study has been performed that  systematically explores the settling mechanism of particles with a non-vanishing Stokes velocity in  turbulent, thermally-driven convection.

Differentiation of a cooling magma is a competitive process between generation, sedimentation and re-entrainment of crystals. The problem of re-entrainment, in particular, has been addressed by various authors both theoretically \citep[e.g.,][]{Solomatov1993b} and experimentally \citep[][]{Solomatov1993b,Lavorel2009}. Although the lifting of negatively buoyant particles from the crests of dunes has been recognized as one of the main mechanisms to keep these in suspension \citep{Solomatov1993b}, in this study we focus entirely on the settling process and completely neglect re-entrainment, which we plan to address in future work. As soon as our particles reach the bottom boundary of the domain (resp.~the top boundary for light particles), we eliminate them from the flow, not allowing any accumulation or subsequent lifting of the sedimented material. 

We use a modeling approach based on a Eulerian-Lagrangian description of the fluid flow and the particulate phase respectively, and track individual trajectory of each particle. Our approach thus captures how the exact flow structure affects the particle motion. This level of detail is still challenging from the experimental point of view and it brings new results when compared to the one-dimensional turbulent diffusion theory. For example, horizontal variations in the distribution of sedimented particles can be evaluated and linked to the large scale circulation of the fluid (also called the ``wind of turbulence'', see e.g.~\citet{Ahlers2009} for a discussion).

Based on the experimental work of \citet{Koyaguchi1990}, \citet{Sparks1993} argued for cyclic sedimentation of crystals in magma chambers caused by the cessation of convection due to the particle concentration exceeding a certain critical value. Similar behaviour was observed by \citet{Hoeink2006} in the context of numerical simulations of metal-silicate separation, and by \citet{Verhoeven2009} in a numerical study that combines a finite volume convection code with a discrete element method \citep{Cundall1979}. Here we neglect the influence of particles on the convective flow \citep[e.g.][]{Park2018}, i.e.~we assume only small particle concentrations. We believe that the dynamics of dilute suspensions is sufficiently rich to warrant a study entirely dedicated to particle settling before considering additional complexities arising from larger solid fractions.

\rev{For particles with small radii \citet{Verhoeven2009} obtain statistically stationary suspension in which convective motions keep particles indefinitely entrained. Similar results were obtained in the non-rotating cases of \citet{Maas2015, Maas2019} whose model builds on the one by \citet{Verhoeven2009}. Although it is at odds with the early results of \citet{Martin1989}, \citet{Maas2019} conclude that ``it is generally assumed that vigorous convection would prevent major gravitational segregation in a magma ocean at all latitudes (Andrault et al., 2017)''. Here we employ a more elaborate particle model and refute such statement, confirming the experimental results of \citet{Martin1989} in which particles with a vanishing response time settle at an exponential rate.}

The rest of the paper is organized as follows. In Section \ref{sec:GovEq} we introduce our numerical model and discuss the choice of model parameters. In Section \ref{sec:res} we present the settling curves of a reference simulation and classify them according to four distinct regimes. In Section \ref{sec:Poisson} we then introduce a general model that describes particle settling as a random process. In Section \ref{sec:horizontal} we discuss the horizontal distribution of sedimented particles, showing how it can be strongly non-uniform in some regimes. In Section \ref{sec:beta} the focus is on particles lighter than the fluid, including bubbles. These become concentrated in flow vortices, which significantly delays their rising. In Section \ref{sec:Ra} we analyze how our results depend on the strength of convective vigor and fluid inertia. Finally, in Section \ref{sec:MO} the results are extrapolated to the environment of an extremely vigorous, global magma ocean and that of a large magma chamber.

\section{Governing equations}\label{sec:GovEq}
\noindent
Rayleigh-B\'enard convection of an incompressible isoviscous fluid is governed by the Boussinesq equations:
\begin{eqnarray}\label{NVSeq1}
    \partial_\tau \mathbf{U} + (\mathbf{U} \cdot \mathbf{\nabla})\mathbf{U} &=& -\mathbf{\nabla} P /\rho_0 + \nu \nabla^2 \mathbf{U} - \alpha (T-T_0) \mathbf{g},\\
    \mathbf{\nabla} \cdot \mathbf{U} &=& 0,\\
    \partial_\tau T + (\mathbf{U} \cdot \mathbf{\nabla})T &=& \kappa \nabla^2 T,
    \label{NVSeq3}
\end{eqnarray}
where $\mathbf{U}(\mathbf{X},\tau)$ and $T(\mathbf{X},\tau)$ are respectively the velocity and temperature fields, $\nu$ the kinematic viscosity of the fluid, $\rho_0$ the mean mass density at the reference temperature $T_0$, $\alpha$ the volumetric thermal expansion coefficient with respect to the reference temperature, $\bm{g}$ the gravitational acceleration, $\kappa$ the thermal diffusivity. The hydrostatic stress, $\nabla P_0 = \rho_0 \mathbf{g}$, is already subtracted from Eq.~\eqref{NVSeq1}, leaving only the dynamic pressure $P$ on its right-hand side (RHS). As shown by the last term in Eq.~\eqref{NVSeq1}, only temperature-induced variations of density are considered to drive the flow.

Equations \eqref{NVSeq1} -- \eqref{NVSeq3} are solved in a 2D box with periodic side walls and aspect ratio 2. No-slip conditions are assumed on the top and bottom boundaries, which are are isothermal, with a constant temperature difference $\Delta T$ driving thermal convection.

We non-dimensionalize the governing equations by scaling the length with the height of the box $H$, $\mathbf{x} := \mathbf{X}/H$, the velocity with the characteristic velocity $u^* := \sqrt{\alpha g \Delta  T H}$, $\mathbf{u}:=\mathbf{U}/u^*$, the density with the reference density $\rho_0$, and we introduce non-dimensional temperature $\theta := (T-T_0)/\Delta T$. For the time and pressure it then follows: $t:=\tau u^* / H$, and $p:=P/(\rho_0  u^{*2})$.

In terms of non-dimensional quantities, the governing equations read:
\begin{eqnarray}\label{NVSnonD1}
    \partial_{t} \mathbf{u} + (\mathbf{u} \cdot \mathbf{\nabla})\mathbf{u} &=& -\mathbf{\nabla} p  + \sqrt{\frac{Pr}{Ra}} \nabla^2 \mathbf{u} + \theta\, \mathbf{\hat{z}}\\
    \mathbf{\nabla} \cdot \mathbf{u} &=& 0\\
    \partial_{\tau} \theta + (\mathbf{u} \cdot \mathbf{\nabla})\theta &=& \frac{1}{\sqrt{Pr Ra}} \nabla^2 \theta ,
    \label{NVSnonD3}
\end{eqnarray}
where $Ra$ and $Pr$ are the Rayleigh and Prandtl number that control the flow characteristics: %
\begin{equation}
Ra: = \frac{\alpha g \Delta T H^3 }{\nu \kappa}, \quad Pr:=\frac{\nu}{\kappa}.
\end{equation} 

The fluid carries inertial particles, whose trajectory is governed by friction from the surrounding fluid in combination with particle buoyancy. Under idealized conditions of spherically-shaped particles with small Reynolds number, the Lagrangian equation of motion for a massive particle reads \citep[e.g.][]{Mathai2016}:
\begin{equation}\label{eqMR}
\der{\mathbf{V}}{\tau} = \beta \frac{D\mathbf{U}}{Dt} + \frac{1}{\tau_D}(\mathbf{U}-\mathbf{V}) + (1 - \beta )\mathbf{g},
\end{equation}
where $\mathbf{V}$ is the particle velocity and the first term on the RHS denotes the material derivative of the fluid velocity. The modified density ratio $\beta = 3\rho_\mathrm{f} /(\rho_\mathrm{f} + 2\rho_\mathrm{p})$ relates the density of the fluid $\rho_\mathrm{f}$ with the particle density $\rho_\mathrm{p}$. Eq.~\eqref{eqMR} is a truncated version of the original equation derived independently by \citet{Maxey1983} and \citet{Gatignol1983}. Due to the small Reynolds number and size of the particles we neglect here both the unsteady drag term, known as history term, and the Fax\'en corrections.  

We systematically vary the ratio $\rho_\mathrm{f}/\rho_\mathrm{p}$ and assume that it is constant for each particle throughout the simulation. This assumption does not necessarily neglect the density variations of material: in view of the Boussinesq approximation $\rho_\mathrm{f}=\rho_0[1-\alpha(T-T_0)]$, employed in Eq.~\eqref{NVSeq1}, the assumption simply means that each particle is always at the same temperature as the surrounding fluid and has the same thermal expansivity \citep[for effects resulting from keeping the particles at a different temperature than the fluid, see][]{gan2003}. The particle response time $\tau_D = \rc^2 / (3 \nu \beta)$ depends quadratically on the particle radius $\rc$, which we also vary systematically.

After non-dimensionalizing it with the same scales introduced above, Eq.~\eqref{eqMR} takes the form:
\begin{align}
\der{\mathbf{v}}{t}=\beta \frac{D\mathbf{u}}{Dt} + \frac{1}{\St}(\mathbf{u}-\mathbf{v}) + \Lambda \,\mathbf{\hat{z}}, \label{eqnonDMR}
\end{align}
leaving three non-dimensional parameters to control the particle dynamics, namely:
\begin{equation}\label{eqStLBeta}
\beta= \frac{3\rho_\mathrm{f} }{\rho_\mathrm{f} + 2\rho_\mathrm{p}}; \quad 
\St = \frac{\rc^2 \sqrt{\alpha g \Delta  T}}{3 \nu \beta \sqrt{H}}; \quad 
\Lambda = \frac{\beta{-}1}{\alpha \Delta T}.  
\end{equation}
The Stokes number $\St$ is the particle response time $\tau_D$ divided by the characteristic time $H/u^*$. It describes the viscous friction acting on each particle due to the difference between particle and fluid velocity. The parameter $\Lambda$ (hereafter buoyancy ratio) expresses the relative importance of particle buoyancy with respect to the thermally-induced buoyancy of the fluid (the unit vector $\hat{\bf{z}}$ points vertically upward). The first term on the RHS of Eq.~\eqref{eqnonDMR} is the so called added-mass as estimated by \citep{Auton1987}. We do not consider any feedback mechanism with respect to the flow: the fluid velocity $\bm{u}$ is obtained from Eqs.~\eqref{NVSnonD1}--\eqref{NVSnonD3} and does not depend on the particle velocity $\bm{v}$, i.e.~we adopt a one-way coupling \citep[for the distinction between one-way and two-way coupling, see the review of][]{Balachandar2010}.

In a turbulent flow, the adopted particle model can be considered appropriate as long as the particle size, $\rc$ (resp.~$\rc/H$ in dimensionless units) is up to the same order of magnitude as the spatial dissipative scale of turbulence, $\eta$. In Rayleigh-B\'{e}nard flow the global value of such scale, in the current dimensionless units, goes as $\eta = Pr^{1/2}(Ra(Nu-1))^{-1/4}$ \citep[e.g.][]{Shraiman1990}, meaning that it decreases at increasing the thermal forcing $Ra$ (and so the Nusselt number $Nu$) but increases at increasing the Prandtl number $Pr$ (see also Discussion).

For particles suspended in a fluid at rest, i.e.~with $\bm{u}\equiv 0$, Eq. \eqref{eqnonDMR} can be solved analytically, yielding:
\begin{equation}\label{eqrest}
\tilde{\bm{v}}(\bm{x},t) = \bm{v}_0(\bm{x}) \,\exp\left(\frac{-t}{\St}\right) + \St \, \Lambda \,\hat{\bm{z}}.
\end{equation}
In the limit $t{\rightarrow}\infty$, the so-called terminal or Stokes' velocity $\vt$ is reached:
\begin{equation}\label{eqvt}
\tilde{\bm{v}}(t{\rightarrow}\infty) =  \St \, \Lambda \,\hat{\bm{z}} = \frac{2}{9}\,\frac{\rho_\mathrm{f}-\rho_\mathrm{p}}{\nu\rho_\mathrm{f}}\,\frac{\rc^2 g}{u^*}\,\hat{\bm{z}} =: - \vt \,\hat{\bm{z}},
\end{equation}
where the terminal velocity is defined positive for sinking particles and negative for rising particles.

We inject $10^6$ particles of 301 different types into a fully developed, two-dimensional, statistically-steady thermal convection, with each particle type represented by three values: $St$, $\Lambda$, and $\beta$. Since we are primarily interested in the dynamics of the particles, we refer to the thermal flow of the carrier as the ``background'' flow. For particles with $\vt>0$ ($\Leftrightarrow \Lambda<0 \Leftrightarrow \beta<1$), i.e.~those denser than the fluid (labeled as heavy), we measure the time it takes until they reach the bottom boundary. For particles with $\vt<0$ ($\Leftrightarrow \Lambda>0 \Leftrightarrow \beta>1$, labeled as light) we do the same with respect to the top boundary. For brevity, both these cases are referred to as ``settling''. Initially, all particles are distributed uniformly across the domain and their velocity is set equal to the local velocity of the fluid. 300 different types of particles are obtained by evenly sampling $\rho_\mathrm{f}/\rho_\mathrm{p}$ and $\rc^2$; one particle type is reserved for fluid tracers.

The above described model system is numerically simulated by means of the Eulerian-Lagrangian code \textit{ch4-project} \citep{Calzavarini2019}. The code adopts a  lattice Boltzmann (LB) algorithm for the computation of the fluid and temperature dynamics, while it uses a second order time-stepping and grid-to-particle bi-linear interpolation for the computation of particles' trajectories. This code has been already extensively employed in studies involving turbulent thermal convection and inertial particle dynamics \citep{Calzavarini2020}. 

\subsection{Model parameters}\label{sec:modpar}

We aim to map the settling behaviour of particles over the entire $\St, \Lambda$, and $\beta$ space, while focusing on highly vigorous convection ($Ra=[10^{8}, 10^{10}, 10^{12}]$), with moderate to small importance of inertia ($Pr=[10,50]$). As such, our results are applicable to a range of natural systems (see Section \ref{sec:Intro}). Throughout the paper we strictly use non-dimensional control parameters, but it is instructive to demonstrate how these are linked to physical parameters of a particular system, namely the thermal convection of a large reservoir of crystallizing magma. In this section we inspect how the parameter space is mapped and discuss intrinsic limitations of our numerical approach.

In Table \ref{magpar} we list the physical parameters that roughly describe the thermal convection of a primordial, mantle-deep magma ocean for the Earth. A relativley large uncertainty is in the value of the viscosity of high-pressure and -temperature magma. First-principles simulations suggest that the kinematic viscosity of MgSiO$_3$, one of the major mantle silicates, over the temperature and pressure range relevant for a global magma ocean ($\sim 2000-4000$ K and $0-130$ GPa) is on average of the order of $10^{-5}-10^{-6}$ m$^2$/s \citep{Karki2010}. Since the Prandtl number is defined as $\nu/\kappa$, the lower and upper bounds of $\nu$ define the range of interest of $Pr$ and we indicate $\nu$ directly as $Pr\times\kappa$ in Table \ref{magpar}. The temperature contrast $\Delta T$ driving convection is also difficult to determine precisely. The reference value of only 1 K reported in the table roughly corresponds to the contrast predicted by parameterized models of the thermal evolution of the Earth's magma ocean in the presence of an atmosphere \citep[e.g.,][]{Lebrun2013,Nikolaou2019}. Such a low value is also representative for planetary cores, where a large volume of low-viscosity metallic liquid undergoes thermal convection \citep[e.g.,][]{Christensen2010}.

\begin{table}[ht]
\caption{Parameters of a global, mantle-deep magma ocean}
\label{magpar}
\centering
\begin{tabular}{l c c c}
\hline
 Parameter & Symbol &  Value & Units \\
\hline
  Mantle depth  & $H$ & $2890$ & km \\
  Reference grav.~acceleration & $g_\mathrm{ref}$ & $9.8$ & m/s$^2$ \\
  Thermal expansivity$^a$ & $\alpha$ & $5{\times}10^{-5}$  & K$^{-1}$\\
  Thermal diffusivity$^b$ & $\kappa$ & $5{\times}10^{-7}$& m$^2$/s  \\
  Kinematic viscosity$^c$ & $\nu$ & $[10,50]{\times}\kappa$ & m$^2$/s \\
  Temperature contrast$^d$ & $\Delta T$ & 1 & K \\
  Crystal size$^a$ & $\rc^\mathrm{ref}$ & $[0.5, 10]$ & mm \\
  Density ratio & $\rho_\mathrm{p}/\rho_\mathrm{f}$ & $[0, 2]$ & -- \\
\hline
\end{tabular}
\\ $^a$ from \citet{Solomatov2015}; $^b$ from \citet{Huaiwei2015}; $^c$ see  \citet{Karki2010} for typical viscosities of silicate liquids at high pressure and temperature; $^d$ see e.g. \citet{Lebrun2013} and \citet{Nikolaou2019} for typical temperature contrasts during the evolution of magma oceans.
\end{table}

Sampling the ranges of $\rc^\mathrm{ref}$ and $\rho_\mathrm{p}/\rho_\mathrm{f}$ from Table \ref{magpar} results in a sampling of the non-dimensional parameter space $\beta, \St, \Lambda$ (set A in Fig.~\ref{figStLspace}). The y-axis in Fig.~\ref{figStLspace} represents the absolute value $|\Lambda|$ rather than $\Lambda$ in order to fit both light ($\Lambda>0$) and heavy ($\Lambda<0$) particles into a compact plot. The modified density ratio $\beta$ is marked by color only. Later we will show that, apart from the effect described in Section \ref{sec:beta}, the first term on the RHS of Eq.~\eqref{eqnonDMR} has secondary importance on the settling behaviour, which sidelines the relevance of $\beta$.

\begin{figure}[ht]
\includegraphics[scale=0.3]{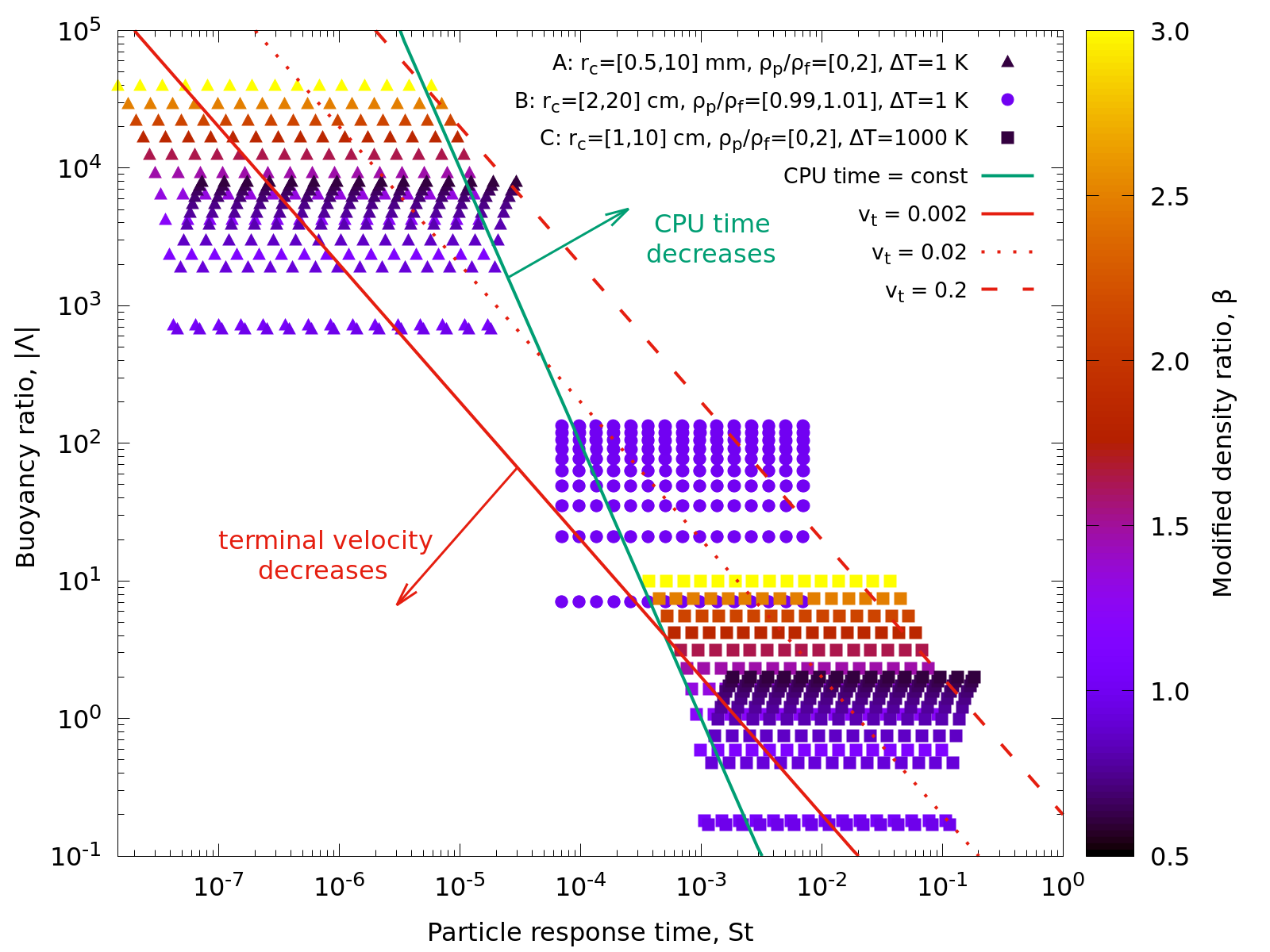}
\caption{Model parameter space describing particle dynamics. Red dashed and solid lines show three isolines of the terminal velocity $\vt$. Colored symbols show three different simulation sets: A, B, and C (triangles, circles, and squares). The green line marks a constant value of required CPU time. We performed numerical simulations corresponding only to the sets B and C (see text for details). $Pr$ is equal to 50.}
\label{figStLspace}
\end{figure}

The terminal velocity $\vt$ and the response time $\St$ can be used to a priori estimate the number of time steps that are required to evaluate the settling time of a given particle type. In the time $1/\avt$, each particle would cross the model domain vertically if sinking (or rising) at the speed $\vt$, making $1/\avt$ a proxy for the minimum required duration of a simulation (and thus CPU time). Due to constraints arising from the numerical integration of Eq.~\eqref{eqnonDMR}, which we explain below, we use max$(10 \Delta t/St,1) / \avt$ to estimate the minimum required CPU time (green line in Fig.~\ref{figStLspace}). Here, $\Delta t$ denotes the maximum time step allowed by the Courant-Friedrichs-Lewy condition, i.e.~by the advective and diffusive time scales of the background flow.

For illustration purposes, let us assume $\bm{u}\equiv 0$ and discretize Eq.~\eqref{eqnonDMR} via explicit, first-order Euler scheme:
\begin{equation}\label{eqtaudproblem}
\tilde{\bm{v}}_{\mathrm{n}+1} = \tilde{\bm{v}}_\mathrm{n} \left( 1-\frac{\Delta t}{\St}\right) + \frac{\Delta t}{\St} \bm{v}_\mathrm{t}.
\end{equation}
It follows that ${\Delta t}/{\St}$ must be smaller than 2 in order to avoid numerically unstable solutions. Demanding numerical accuracy limits the admissible values of ${\Delta t}/{\St}$ even further -- only a small fraction of $\bm{v}_\mathrm{t}$ must be added at each time step to ensure that convergence to the stationary solution $\bm{v}_\mathrm{t}$ is smooth. In a turbulent flow ($\bm{u}\neq 0$), Eq.~\eqref{eqnonDMR} yields accurate trajectories only when ${\Delta t}/{\St}<0.1$ (the exact value depends on the employed numerical scheme, with 0.1 resulting from our experience with the second-order Adams-Bashforth formula that we use to advect the particles). This constraint increases the CPU time of each simulation (resp.~decreases the allowed time step) by an additional factor, $10 \Delta t/St$, where $St$ is the smallest Stokes number in the respective set of particles.

Red lines in Fig.~\ref{figStLspace} mark isolines of $\vt$. In Section \ref{sec:res}, we show that to first-order the settling behaviour of particles can be described by their terminal velocity only: for a given background flow, particles with the same $\vt$ settle in a similar manner. Since the green and red lines in Fig.~\ref{figStLspace} have different slopes, it is convenient to modify the parameters from Table \ref{magpar} to move along the isolines of $\vt$ in the direction of smaller CPU time (i.e.~to the right of the $\St, |\Lambda|$ space). Sets B and C in Fig.~\ref{figStLspace} are two such modifications of the original set A, obtained by setting: B) $\rc^\mathrm{ref} = \langle 2, 20\rangle$ cm, $\rho_\mathrm{p}/\rho_\mathrm{f}=\langle 0.99, 1.01\rangle$, and C) $\rc^\mathrm{ref} = \langle 1, 10\rangle$ cm, $\rho_\mathrm{p}/\rho_\mathrm{f}=\langle 0, 2\rangle$, $\alpha=2\times 10^{-4}$, $\Delta T=1000$ K. While set A is computationally difficult to reach and would require close to a year on several hundreds of CPU cores, sets B and C can be completed within a month on a 32-core machine.

For the purpose of this study, particle sets B and C can be simply understood as the selected coverage of model parameter space (later we argue that based on these sets we map the entire $St, \Lambda, \beta$ space reasonably well). We note, however, that both sets also have a certain geophysical interpretation. When compared to the original set A, set B has enlarged $r_\mathrm{c}$ and a narrowed $\rho_\mathrm{p}/\rho_\mathrm{f}$ range. As such, its parameters roughly correspond to large clusters of crystals with density similar to that of the surrounding magma. Set C has a reduced thermal expansivity and a larger temperature contrast that lies in the range of temperature contrasts characteristic of magma oceans that cool in absence of an atmosphere \citep{Lebrun2013,Nikolaou2019}. 

Based on the parameter values listed in Table \ref{magpar}, the Rayleigh number $Ra_\mathrm{ref}$ of a mantle-deep magma ocean would be of the order of $10^{27}$. This is far from being reachable with any numerical method because the thickness of the thermal boundary layer in a convecting system scales approximately as $Ra^{-1/3}$, demanding higher resolution for higher $Ra$. Here we model a series of Rayleigh numbers up to $Ra=10^{12}$, using up to $4096{\times}2048$ grid points. The possibility of extrapolating our results to higher Rayleigh numbers is analyzed in Section \ref{sec:MO}. 

We note that our code is based on a dimensional formulation. Therefore, in order to reduce the Rayleigh number we need to modify some of the parameters in Table \ref{magpar}. Our aim is to modify the parameters such as to change $Ra$ and leave the remaining control parameters $Pr,\, St,\,\Lambda$, and $\beta$ untouched, regardless of the choice of $Ra$. This can be achieved by replacing $g_\mathrm{ref}$ with a reduced gravitational acceleration, $g:=g_\mathrm{ref} Ra / Ra_\mathrm{ref}= \nu\kappa Ra / (\alpha \Delta T H^3)$, and by replacing $\rc^\mathrm{ref}$ with an inflated crystal size, $\rc:=\rc^\mathrm{ref} (g_\mathrm{ref}/g)^{1/4}$. In this way, the coverage of the $St,\,\Lambda,\,\beta$ space remains identical for all tested values of $Ra$ (i.e.~Fig.~\ref{figStLspace} remains the same regardless of the value of $Ra$). 

For each simulation set, we first wait for thermal convection to develop into a statistically steady state and then we inject all the particles at once, distributing them uniformly in space and assigning them the velocity of the carrier fluid, i.e. $\bm{v}(t{=}0,x,z)=\bm{u}(t{=}0,x,z)$. In Fig.~\ref{figvrms}a, we show the average root mean square velocity of the background flow for all the tested values of $Ra$ and $Pr$. Fig.~\ref{figvrms}b shows the corresponding Reynolds number, $Re:=U_\mathrm{rms}\,H/\nu$. For $Pr=10$ we run simulations with $Ra$ equal to $10^{8}$ and $10^{10}$, while for $Pr=50$ we test three values, $Ra=10^{8}$, $10^{10}$, and $10^{12}$ (lowering Prandtl number increases the resolution demands -- see the Reynolds number for two simulations with the same $Ra$ but different $Pr$). Our simulation sets are labeled as B or C, depending on the range of particle parameters (see Fig.~\ref{figStLspace}), and by upper and lower indices we label the exponent of $Ra$ and the value of $Pr$ (e.g.,~C$^{10}_{50}$ stands for simulation set C with $Ra=10^{10}$ and $Pr=50$).

\begin{figure}[ht]
\includegraphics[scale=0.33]{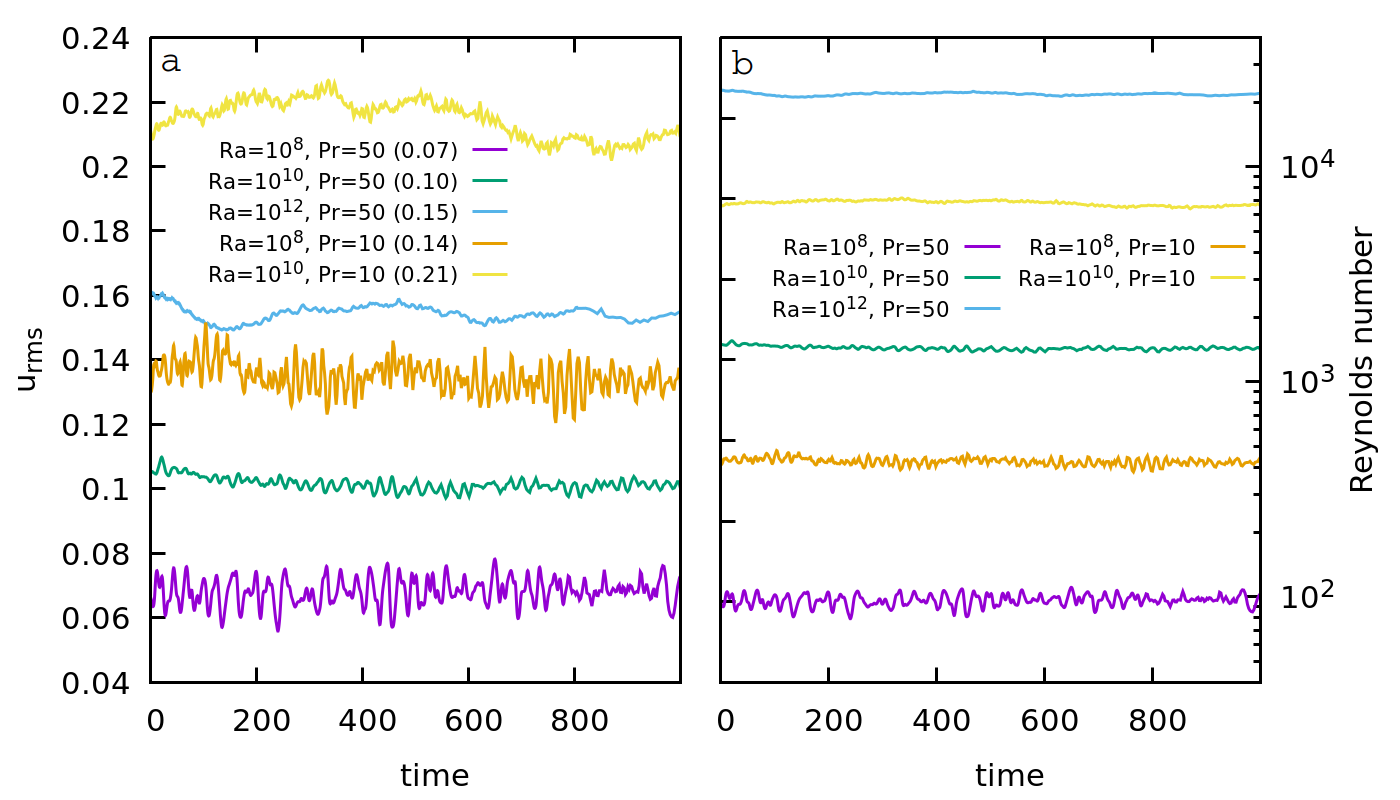}
\caption{(a) Volumetric averages of the root mean square velocities in simulation sets C$^{8}_{50}$, C$^{10}_{50}$, C$^{12}_{50}$, C$^{8}_{10}$, and C$^{10}_{10}$. The number in parenthesis indicates the time-averaged value $\vm$ that is used later for computing the $\vt/\vm$ ratio. (b) The corresponding Reynolds number, $U_\mathrm{rms}\,H/\nu$.}
\label{figvrms}
\end{figure}

\section{Results: settling curves}\label{sec:res}

In Section \ref{sec:modpar} we anticipated that the terminal velocity $\vt$ is capable of sorting the settling behaviour of particles in a flow of given $Ra$ and $Pr$. In order to compare the settling behaviour also across flows with a different convective vigor, one more parameter is needed. Similarly to the experimental study of \citet{Martin1989}, who divide $\vt$ by the average vertical velocity of the flow and use the resulting ratio to organize their results, we use $\vm$ to account for the properties of the background flow. In this section, we show that there are four distinct regimes describing the sinking or rising of particles, and that the ratio $\avt/\vm$ determines to which regime a given particle type belongs. 

In Fig.~\ref{fig4stagioni} we plot the temporal evolution of the settling process for the simulation set C$^{10}_{50}$, which is taken as a reference case. When heavy particles ($\vt{>}0$) reach the bottom, resp.~light particles ($\vt{<}0$) reach the top, we eliminate them from the flow and mark them as settled. Each line in Fig.~\ref{fig4stagioni} represents a different particle type, although we do not show the respective values of $St, \, \Lambda,$ and $\beta$. Instead, we mark each line by the value of $\avt/\vm$ -- this single parameter uniquely orders the obtained settling curves. We identify four distinct groups: i) ``stone-like'' regime, $\avt/\vm > 2.0$ (Fig.~\ref{fig4stagioni}a); ii) bi-linear regime, $0.3 < \avt/\vm < 2.0$ (Fig.~\ref{fig4stagioni}b); iii) transitional regime, $0.02 < \avt/\vm < 0.3$ (Fig.~\ref{fig4stagioni}c); and iv) ``dust-like'' regime $\avt/\vm < 0.02$ (Fig.~\ref{fig4stagioni}d).

The time on the x-axis of Fig.~\ref{fig4stagioni} is multiplied by $\vt$ for each settling curve individually. The x-axis thus represents distance rather than time. The ``terminal distance'', $t\,\vt$, corresponds to the distance a particle with a given $\vt$ would travel in a fluid at rest by the time $t$ (i.e.~$t\,\vt=1$ represents sinking with the Stokes velocity through the entire container). This means that, even though the particles in Fig.~\ref{fig4stagioni}d seemingly take only approximately 5 times longer than those in Fig.~\ref{fig4stagioni}a to completely settle, the actual time differs by more than two orders of magnitude because the corresponding value of $\vt$ differs by more than a factor hundred in both subplots. The same applies for different settling curves within each subplot: two settling curves that overlap but have different colors correspond to different settling rates with respect to time $t$. 

\begin{figure}[ht]
\includegraphics[scale=0.17]{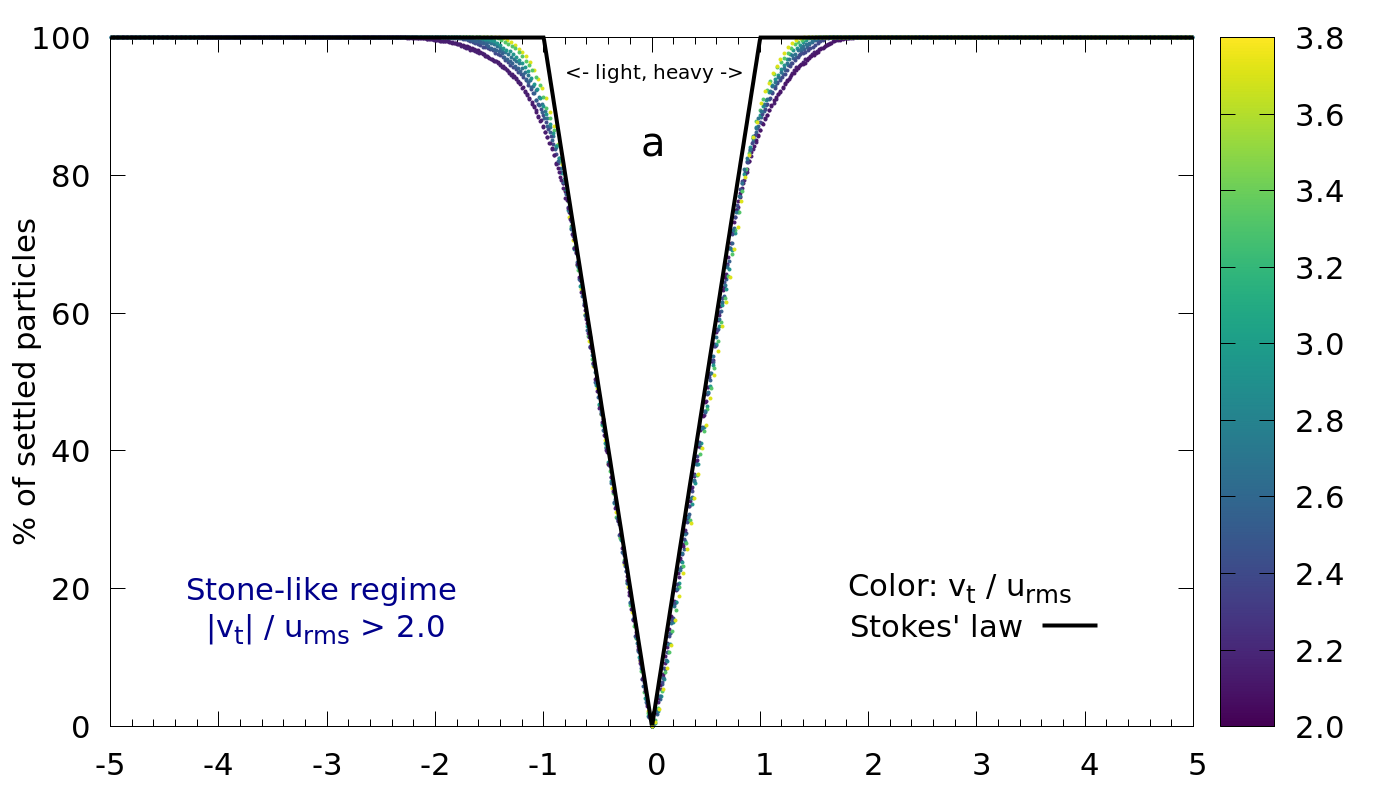}\includegraphics[scale=0.17]{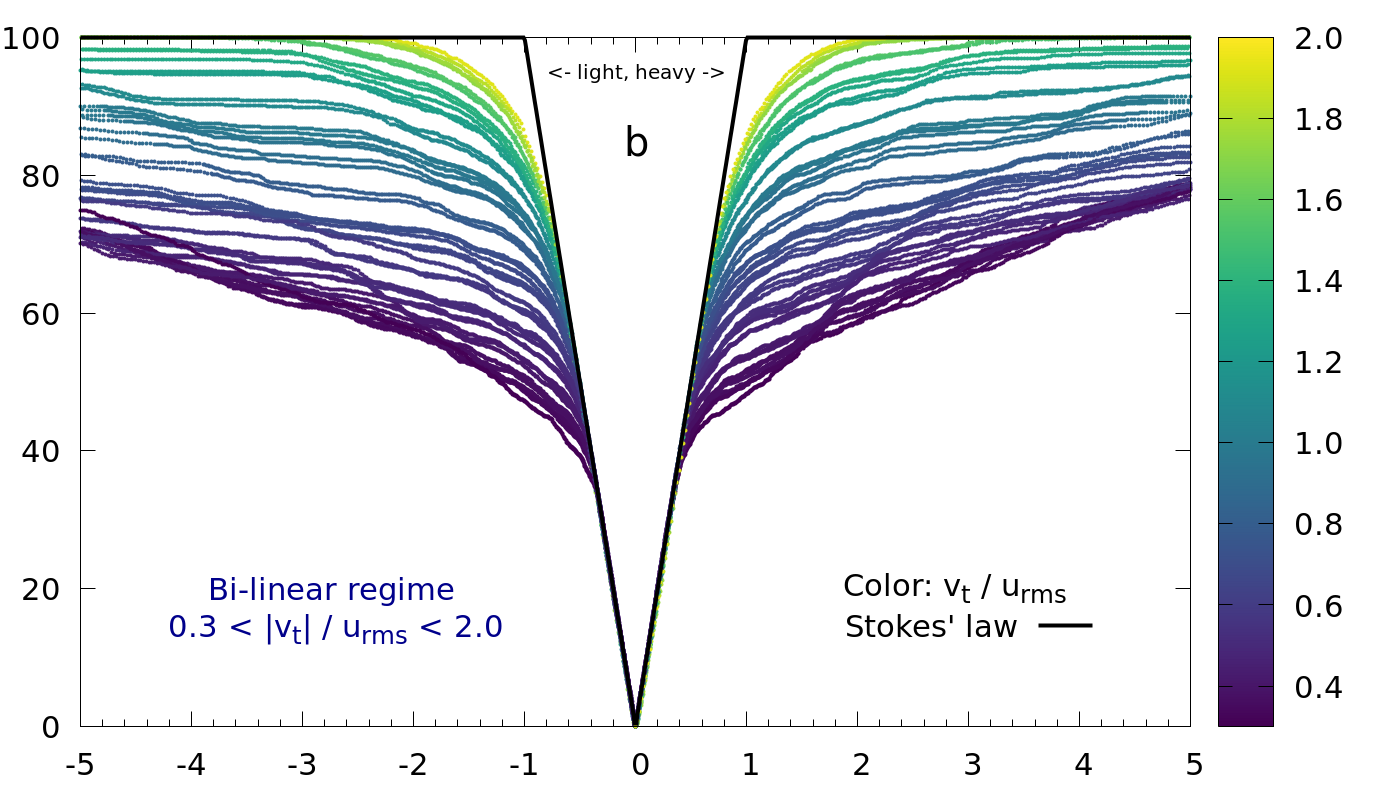}
\includegraphics[scale=0.17]{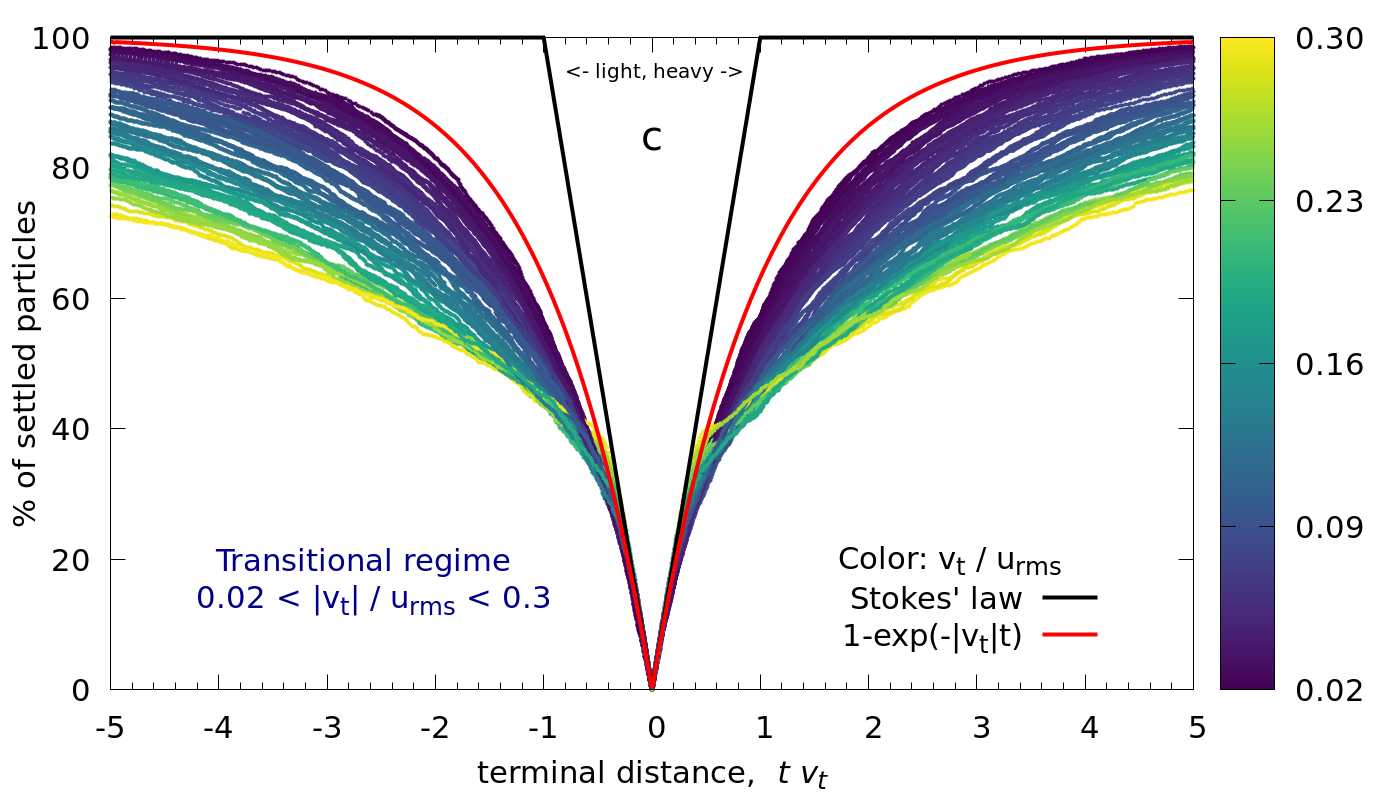}\includegraphics[scale=0.17]{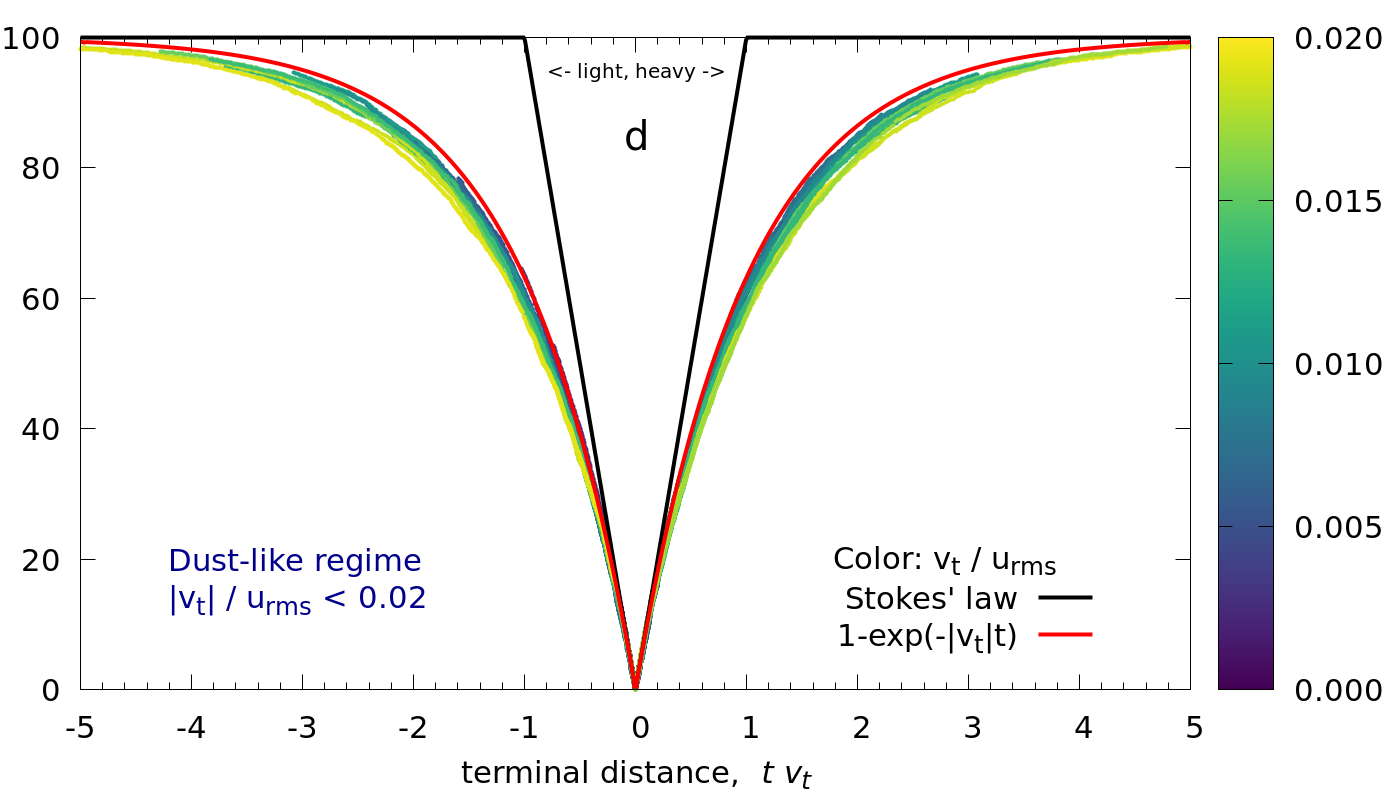} \caption{Settling curves obtained from the simulation set C$^{10}_{50}$. All particle types are separated into four groups based on the $\avt/\vm$ ratio, which is shown in color for each subplot. The x-axis represents the terminal distance $t\,\vt$. Black and red solid lines represent the analytic solutions \eqref{eqrestint} and \eqref{eqNokes} respectively.}
\label{fig4stagioni}
\end{figure}

The black line in Fig.~\ref{fig4stagioni} is the theoretical prediction
\begin{equation}\label{eqrestint}
\frac{N_\mathrm{s}}{N_0} = \int_0^t |\bm{\tilde{v}}|\,\mathrm{d}t' = |\bm{v}_\mathrm{t}| t - |\bm{v}_\mathrm{t}| St \left(1-\exp\left(\frac{-t}{St}\right)\right)
\end{equation}
where $N_\mathrm{s}$ and $N_0$ are the number of settled particles and the initial number of particles respectively. The velocity $\bm{\tilde{v}}$ is given by Eq.~\eqref{eqrest}, and zero initial conditions are considered, $\bm{\tilde{v}}(\bm{x},t{=}0)\equiv \bm{0}$. Eq.~\eqref{eqrestint} thus expresses the percentage of particles that would settle at the time $t$ if initially they were distributed uniformly in a still fluid. The shape of the black curve is nearly identical to simply min$(t\,\vt,1)$ because it takes a negligible time for the particles to accelerate from 0 to $\vt$ (see Section \ref{sec:stone} for a further discussion).

Below we analyze the settling regimes individually and explain underlying mechanisms. In Section \ref{sec:background} we discuss how the regimes' properties and boundaries depend on the characteristics of the background flow.

\subsection{``Stone-like'' regime ($\avt/\vm\gtrsim 2$)}\label{sec:stone}

\noindent
The simplest regime corresponds to the case with a high $\avt/\vm$ ratio. For a particle with $\avt>2\vm$, the average convective velocities are more than twice smaller than the speed at which the particle would be sinking if there was no convection. This implies that particles with this property are little affected by the flow -- they sink almost as if the fluid was at rest because thermal convection is slow relative to the particle's vertical drift.

Therefore, when $\avt/\vm\gtrsim 2$, Eq.~\eqref{eqrest} provides a good prediction of the settling behaviour (Fig.~\ref{fig4stagioni}a). As analyzed later, this result is very robust with respect to the values of $Ra$ and $Pr$ because the background flow is nearly irrelevant in this regime.

For even higher $\avt/\vm$ ratios, the fit to Eq.~\eqref{eqrestint} becomes perfect, and the acceleration from 0 to $\vt$ begins to play a role in the shape of obtained solutions. In Fig.~\ref{figStone} we demonstrate this effect on the simulation set xC$^{10}_{50}$, constructed using particles with $10$ times larger radii than in the reference set C$^{10}_{50}$. For clarity of the figure, we only show a few particle types from the set, with the $\avt/\vm$ ratio ranging from 2 to 50. As $\avt$ and $St$ increase in value, the analytic solution \eqref{eqrestint} loses its symmetry with respect to light and heavy particles because its second term gains in relative importance. The second term in Eq.~\eqref{eqrestint} is not symmetrical with respect to $\vt$: particles with the same terminal velocity but different modified density ratio $\beta$ have different values of $St$ (recall the definitions of $St$, $\Lambda$, and that $\vt:=-St \Lambda$). As a result, light particles ($\beta{>}1$) have a shorter response time $St$ when compared to heavy ones ($\beta{<}1$), and accelerate to $\vt$ faster (cf.~the last term in Eq.~\ref{eqrest}).

\begin{figure}[ht]
\includegraphics[scale=0.33]{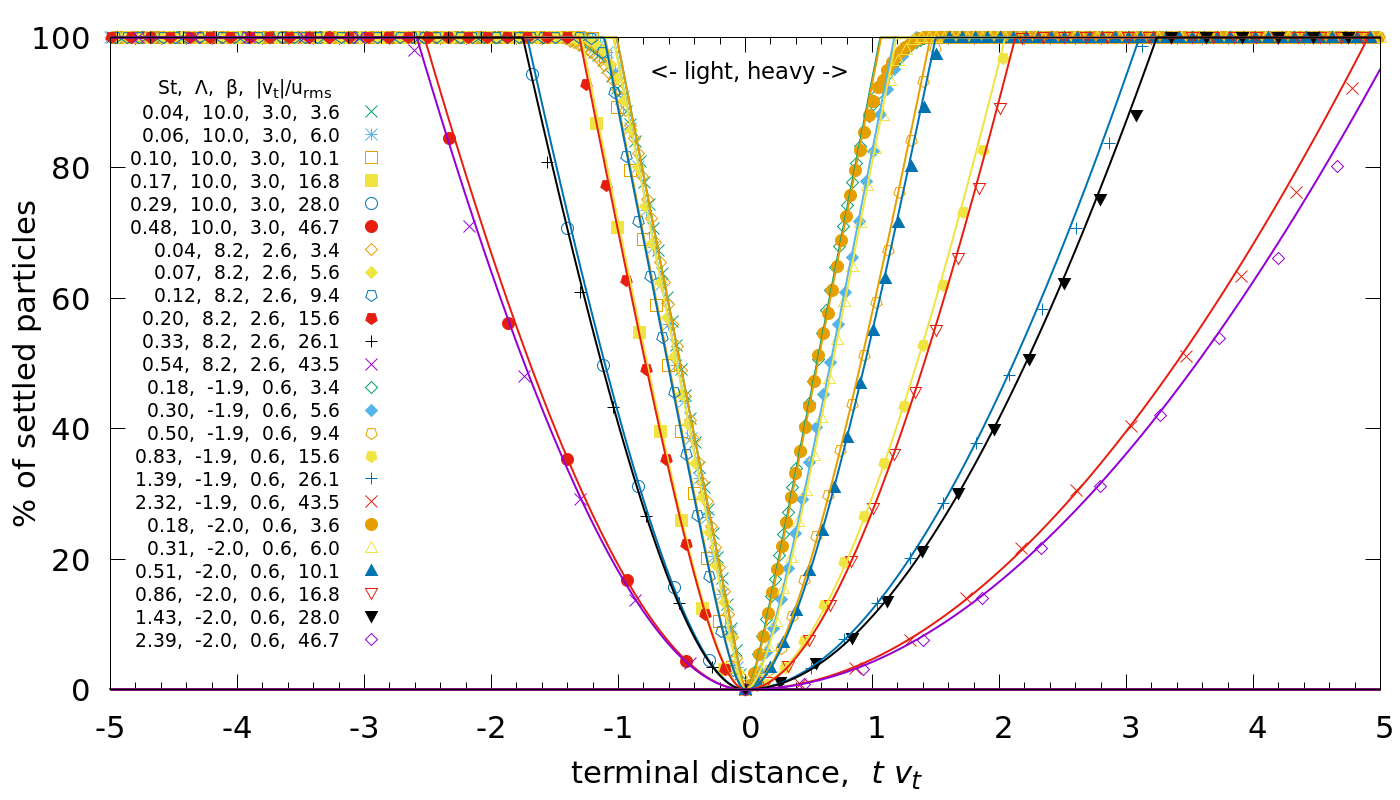}
\caption{Selected settling curves from the simulation set xC$^{10}_{50}$ that consists of particles with a high $\avt/\vm$ ratio (2 to 50 for the depicted selection). Solid lines show the theoretical prediction \eqref{eqrestint}, while colored points are obtained from the numerical simulation.}
\label{figStone}
\end{figure}

We label this regime stone-like. Although particles with $\avt/\vm\gtrsim 2$ still interact with the structure of the flow (see Section \ref{sec:horizontal} below), their vertical speed is close to the free-fall speed $|\vt|$.

\subsection{Bi-linear regime ($0.3\lesssim \avt/\vm\lesssim 2.0$)}\label{sec:Bilinear}
\noindent
Moving to lower $\avt/\vm$ ratios, the settling curves become approximately piecewise linear, with two distinct settling rates. The two different rates correspond to different initial positions of the particles. 

In Fig.~\ref{figBilinear} we depict particles with $0.3<\avt/\vm<2.0$ that are still suspended at the time $t=0.4/(0.3\,\vm)\approx 13$. This snapshot corresponds to the time at which the settling curves of particles with $\avt/\vm = 0.3$ change their slope (cf.~the darkest settling curve in Fig.~\ref{fig4stagioni}b). The particles in Fig.~\ref{figBilinear} form a cloud, centered above a cluster of upwellings, and the larger the value of $\vt$, the smaller is the respective cloud. The surroundings of major downwellings are free of particles. Note that we depict heavy particles only; light particles are located above the central cluster of downwellings.

\begin{figure}[ht]
\includegraphics[scale=0.55]{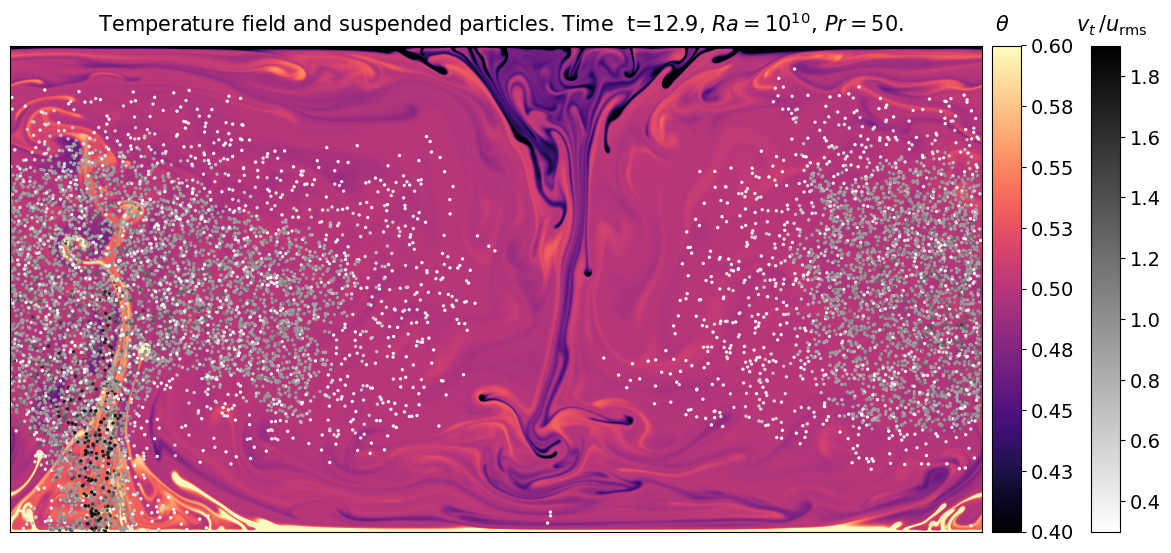}
\caption{Temperature field in the simulation C$^{10}_{50}$ at the time $t=12.9$. The temperature range is clipped for a better visibility of the up- and downwellings. Dots show particles with $0.3<\vt/\vm<2.0$ that have not settled by the respective time. See Supplementary material, Video S1.}
\label{figBilinear}
\end{figure}

After $t\approx 13$, all particles with $0.3<\avt/\vm<2.0$ settle at a reduced rate, while up to this time their settling is well captured by Stokes' law (Fig.~\ref{fig4stagioni}b). It follows that if a heavy particle (with $0.3<\avt/\vm<2.0$) is initially injected close to a major downwelling, or below the top boundary layer where horizontal velocities are large, it settles quickly. Perhaps surprisingly, the settling rate of such particles is represented well by Stokes' law and does not exceed it (compare the first linear segment of all settling  curves in Fig.~\ref{fig4stagioni}b). One could expect the downwellings to sediment the carried particles downstream, speeding up their settling beyond the rate predicted by Stokes' law. This, however, does not happen. In the first stage of the bi-linear regime, particles touch both horizontal boundaries with little to no lateral preference (i.e.~the x-coordinates of the settling events have a uniform distribution), and the percentage of settled particles grows linearly in time, with a slope that matches the Stokes velocity. We pay further attention to the horizontal distribution of settled particles in Section \ref{sec:horizontal}.

In the second stage of the bi-linear regime, the settling rates are significantly reduced. This is because bursts of upwelling flow act against the particles' tendency to settle. The terminal velocities studied in this section are still sufficiently large ($\avt>0.3\,\vm$) for the particles to efficiently penetrate through the fluid flow, but at the same time the existence of plumes alters particle trajectories significantly, in particular by lifting particles that get caught in strong conduits. The higher the $\avt/\vm$ ratio, the less particles survive after the first-stage settling, and the closer they are to the central axis of the major upwelling structure (compare black and white dots in Fig.~\ref{figBilinear}). As a result, the settling rate is smaller for higher $\avt/\vm$ ratios (Fig.~\ref{fig4stagioni}b). Note that throughout the text we refer to the relative settling rate, i.e.~to the slopes of the settling curves in Fig.~\ref{fig4stagioni}, where the x-axis represents distance rather than time. With respect to time $t$, the settling is generally faster for higher $\avt/\vm$ ratios.

For $\avt/\vm\approx 2$, the settling curves are close to being flat in the second stage of the bi-linear regime. This is in agreement with expectations: the velocities of plume heads are typically close to $2\vm$ and the carried particles thus could, in principle, be indefinitely suspended in a fixed point in space by the action of a stationary plume (cf.~Eq.~\eqref{eqnonDMR} with $\bm{u} \equiv -\bm{v}_\mathrm{t}$). For $Ra=10^{10}$ the flow is highly non-stationary and such situation never occurs, but the idealized scenario helps explaining the very slow settling rates.

\subsection{``Dust-like'' regime ($\avt/\vm \lesssim 0.02$)}\label{sec:dust}
\noindent
For $\avt/\vm\lesssim 0.3$, the settling curves smoothly converge towards a single line (red line in Figs.~\ref{fig4stagioni}c and \ref{fig4stagioni}d). In this section we analyze this limiting case, while the transitional regime ($0.02\lesssim \avt/\vm \lesssim 0.3$) is discussed later.

The settling curves are nearly identical when $\avt/\vm \lesssim 0.02$. The very existence of a limit is a non-trivial result. While one can a priori expect the applicability of Stokes' law when $\avt/\vm\rightarrow\infty$, when $\avt/\vm\rightarrow 0$ the particles should behave as fluid tracers. This can be seen directly by inspecting Eq.~\eqref{eqnonDMR}: for a fixed $\Lambda$, if the terminal velocity tends to zero, the Stokes number $St$ also tends to zero. In the limit $St\rightarrow 0$, the second term on the RHS dominates Eq.~\eqref{eqnonDMR}, yielding $\bm{v}=\bm{u}$ (i.e.~fluid tracers). For this reason, $\tau_D$ is called the response time -- particles quickly adapt to the carrier fluid velocity when $\tau_D$ is small. Fluid tracers, however, never touch the bottom nor the top boundary of the model domain and it is thus a priori unclear how particles with a small $\avt/\vm$ ratio should settle.

A simple theoretical model for small particles (i.e.~with a small Stokes number) was developed by \citet{Martin1989}. They proposed that at the base of the model domain, where convective velocities vanish, all particles are free to settle from the fluid with a speed equal to their terminal velocity $\vt$. Therefore, the rate at which the number of particles in the flow $N$ decreases with time is given by
\begin{equation}\label{eqNokes}
\der{N}{\tau} = -A\, \vt\, u^*\, c(0) \quad \Rightarrow \quad N=N_0 \exp\left(\frac{-\vt\,u^*\, \tau}{H}\right) = N_0 \exp(-\vt\, t),
\end{equation}
where $A$ is the area of the base of the domain, $c(z)$ the horizontally averaged concentration of particles at height $z$ above the bottom boundary, and $N_0$ the initial number of particles. The exponential solution in Eq.~\eqref{eqNokes} is obtained by assuming $c(0)$ to be the current average concentration $N/(AH)$. 

Indeed, Fig.~\ref{fig4stagioni}d shows that $1-\exp(-\vt\, t)$ fits the settling curves well for $\avt/\vm \rightarrow 0$, confirming the theoretical and experimental conclusions of \citet{Martin1989}. The derivation of Eq.~\eqref{eqNokes}, however, is based on a counter-intuitive assumption: particles must be uniformly distributed throughout the entire model domain by convection, and yet there must be a boundary layer with little to no mixing, thick enough for the particles to separate from the fluid and accelerate to $\vt$. Moreover, the concentration of particles in the boundary layer is assumed to be the same as in the bulk of convection, without assessing the mutual transport between the two regions.

In the next section, we focus on the statistics of particle transport between convection cells and boundary layers of the flow. We interpret particle settling as a random process, allowing us to provide a quantitative description of the settling regimes' boundaries, and to explain in detail why the exponential law \eqref{eqNokes} fails for particles with $\avt/\vm\gtrsim 0.02$. Most importantly, our approach serves as a unifying theory, capable of containing all four settling regimes in a new equation that estimates the time required for a complete sedimentation of the particles.

\subsection{Particle settling as a random process}\label{sec:Poisson}

The velocity structure of the simulation C$^{10}_{50}$ is shown in Fig.~\ref{figEyes}. On top of the temperature field, shades of green show regions with $ |\bm{u}| < 0.5\,\vm$. Below, these regions are referred to as the ``low-velocity piles'' (the definition may seem rather arbitrary here -- later we investigate piles defined generally as the regions where $ |\bm{u}| / \vm < \pf$ and vary the pile factor $\pf$). The centres of the large scale convection rolls, which also show small velocities, are not considered as the low-velocity piles (thick green line in Fig.~\ref{figEyes}, explained later). Grey points in Fig.~\ref{figEyes} show suspended particles, this time we select only those with $\avt/\vm < 0.3$. For dust-like terminal velocities ($\avt/\vm < 0.02$), the particles appear uniformly distributed, while for $\avt/\vm \rightarrow 0.3$ the spatial distribution resembles the one from Fig.~\ref{figBilinear}. In this section we explore the rate at which particles enter the low-velocity piles, and analyze how likely it is that a particle enters one but does not settle at its base and returns to fast convection cells instead.

\begin{figure}[ht]
\includegraphics[scale=0.55]{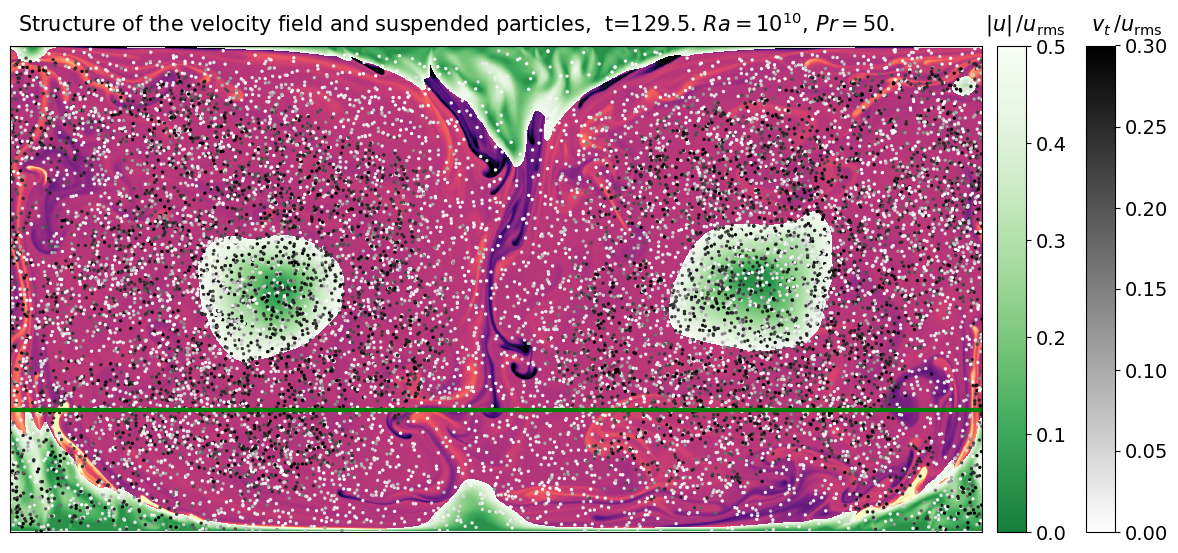}
\caption{Same as Fig.~\ref{figBilinear}, but at time $t=129.5$ and the depicted particles have smaller terminal velocities, $\avt/\vm<0.3$. In green we show regions where thermal convection is slow, with velocities below $0.5 \vm$. The thick green line separates the low-velocity piles from centres of large scale convection rolls, which also show small velocities. See Supplementary material, Video S2.}
\label{figEyes}
\end{figure}

In Fig.~\ref{figCapture} we take a particular instant in time and compare Eq.~\eqref{eqNokes} with the number of particles that have actually settled (red line vs green dots). In agreement with Fig.~\ref{fig4stagioni}, there is a good match with the exponential law for $\avt/\vm \lesssim 0.02$, but for $0.02 \lesssim \avt/\vm \lesssim 1.5$ the settling is slower than Eq.~\eqref{eqNokes} predicts. 

Similarly as in Fig.~\ref{fig4stagioni}, the black line represents $|\vt|t$, i.e.~the prediction based on simple Stokes' settling. Note, however, that while Fig.~\ref{fig4stagioni} maps the temporal evolution of the settling, Fig.~\ref{figCapture} captures only a snapshot in time and does not contain any information about the settling rate. For instance, particles with $\avt/\vm \gtrsim 1.5$ have completely settled at time $t=50$ -- both the exponential and Stokes' laws match the observation and Fig.~\ref{figCapture} cannot be used to distinguish between the two, although Fig.~\ref{fig4stagioni}a shows that particles with a high $\avt/\vm$ ratio follow Stokes' law.

\begin{figure}[ht]
\includegraphics[scale=0.33]{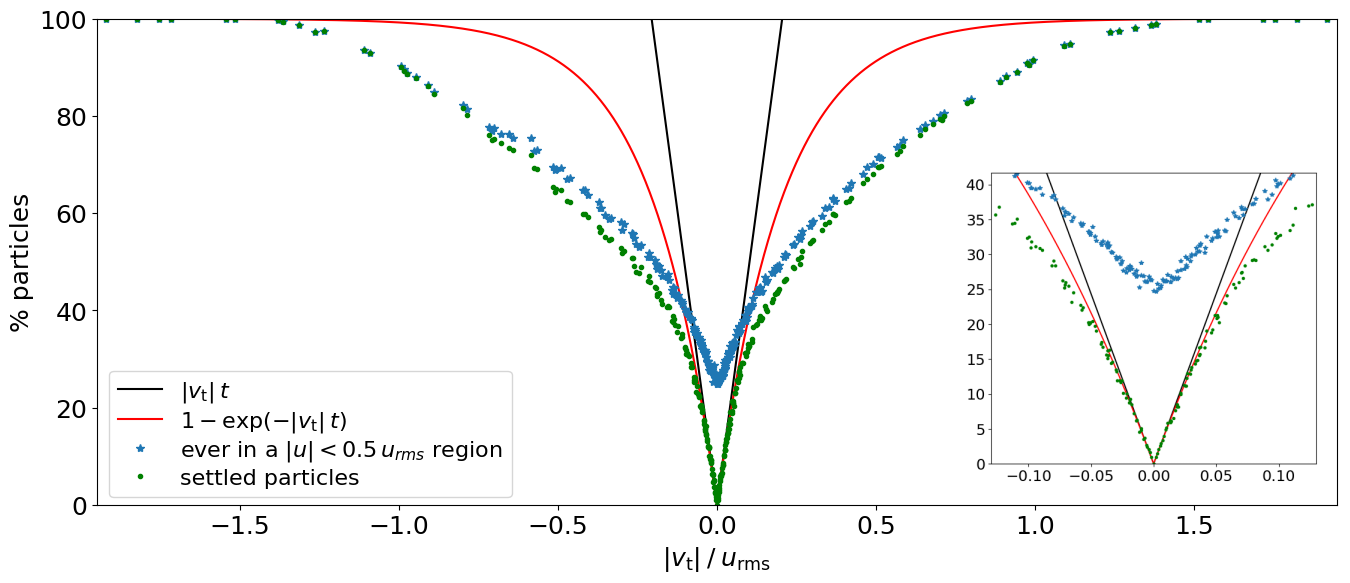}
\caption{Percentage of settled particles (green dots) compared with the two analytic predictions \eqref{eqrestint} and \eqref{eqNokes} (black and red lines, respectively). Blue stars show the percentage of particles that have entered low-velocity piles at least once by the respective time. The figure corresponds to time $t=50$.}
\label{figCapture} 
\end{figure}

Blue stars in Fig.~\ref{figCapture} mark the percentage of particles that have entered the low-velocity piles at least once by the time $t=50$. For heavy particles we only consider the piles located in the bottom fourth of the model domain (thick green line in Fig.~\ref{figEyes}), for light particles only the upper fourth of the domain is considered. The center of the graph ($\vt=0$) represents fluid tracers (treated as heavy for the purpose of this analysis). 

It is important to notice two things: First, the probability of ever entering the low-velocity piles decreases with the $\avt/\vm$ ratio, no matter how small the ratio is. This means that, even in the transitional and dust-like regimes, the trajectories of inertial particles differ from those of fluid tracers: particles governed by the Maxey-Riley equation are more likely to cross sluggish regions of the flow. Second, the probability of escaping the low-velocity piles and returning to the flow increases with the $\avt/\vm$ ratio. This can be seen from the difference between the percentage of particles that have entered the piles and the percentage of particles that have settled: while for higher values of $\avt/\vm$ the two nearly coincide, for lower values they differ, with the difference growing as $\avt/\vm$ approaches zero (compare blue and green symbols in Fig.~\ref{figCapture}). 

\begin{figure}[ht]
\includegraphics[scale=0.33]{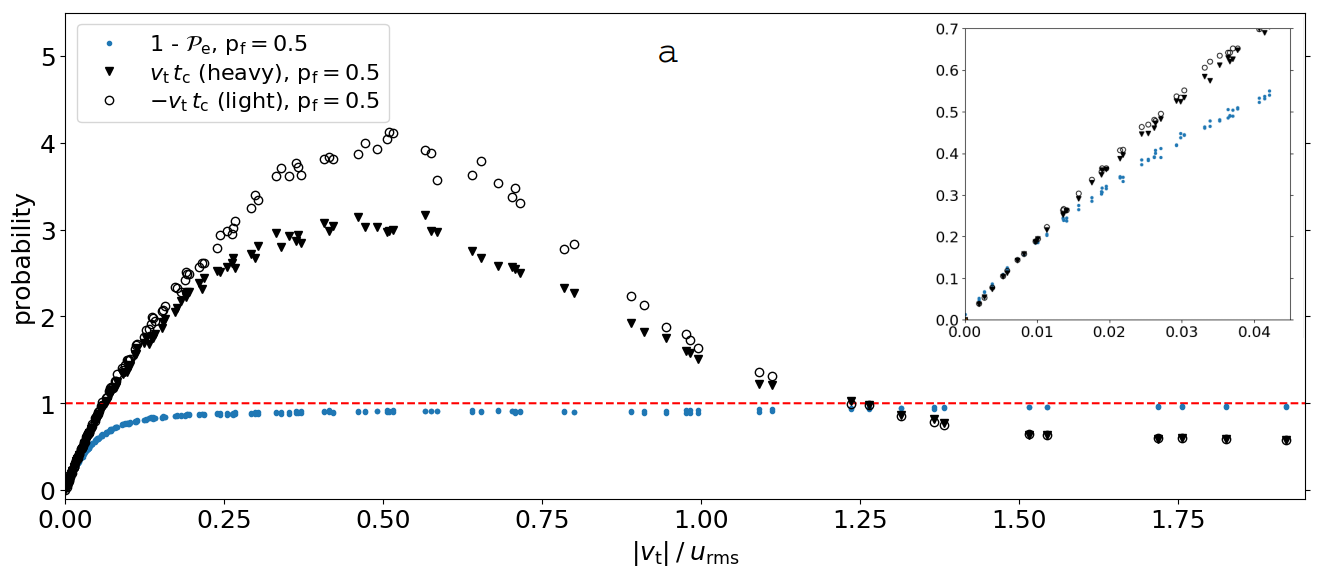}
\includegraphics[scale=0.33]{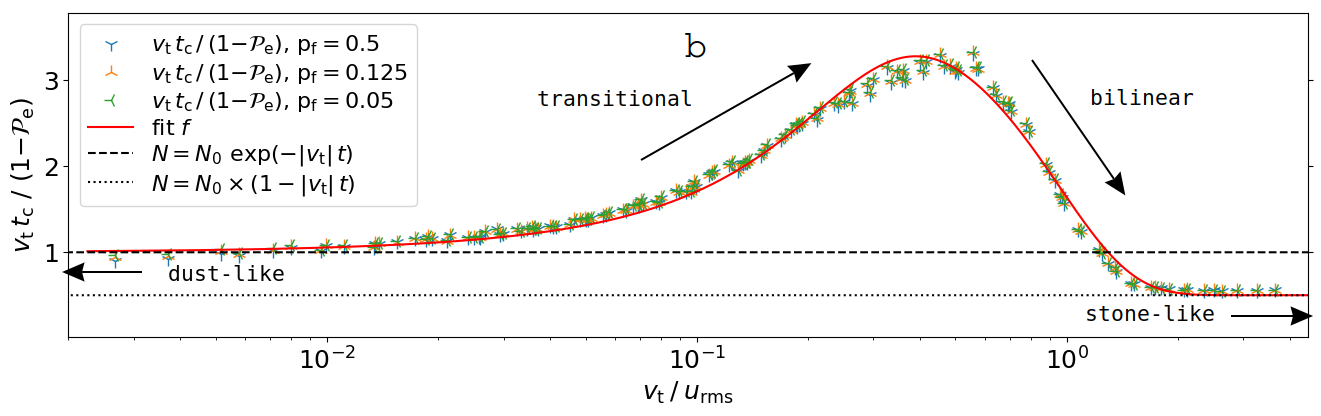}
\caption{(a) Settling probability $1{-}\pe$, where $\pe$ is the probability of escaping from a region with $|\bm{u}|/\vm < \pf$, with $\pf=0.5$. Black symbols show $\vt\,\tc$ (resp.~$-\vt\,\tc$ for light particles), where $\tc$ is the average time between repeated entries into the low-velocity piles. (b) Ratio of $\vt\,\tc / (1{-}\pe)$ for different values of the pile factor $\pf$. The black dashed line shows the reference value 1 because for $\vt\,\tc=(1-\pe)$ the solution \eqref{eqNokes} is obtained. The solid red line shows the skew normal distribution $f$ that is used in our analytic model for particle settling.}
\label{figEscape}
\end{figure}

In Fig.~\ref{figEscape} we plot the probability of escaping the low-velocity piles. It is computed as follows: when a particle reaches a region with $|\bm{u}| < 0.5\,\vm $, we mark it as captured. If a captured particle, instead of settling at the wall of the container, is transported back to a region with $|\bm{u}| > \vm $, we mark it as escaped. Each particle can be captured and escape multiple times and we store the record for each individual particle. The escape probability $\pe$ is simply the sum of all escapes divided by the sum of all captures, taken over all particles of a given type. As expected, $\pe$ goes to one for fluid tracers, because tracers never settle and eventually always escape any low-velocity regions. On the other hand, $\pe$ tends to zero for particles with a large terminal velocity, because such particles always settle upon encountering a slow region.

Now we have the means to describe particle settling as a random process. The exponential law \eqref{eqNokes} and its underlying differential equation $\mathrm{d}N = -\avt N \mathrm{d}t$ are, in a way, calling for such approach: in the terminology of stochastic processes, the equation says that particle settling is a Poisson counting process with intensity $\avt$. 

A random process is defined through an event of a given probability. Here, the event is the settling to the base of a low-velocity pile, with the probability equal to $1{-}\pe$. For each particle type, the number of particles that settle over a time  $\mathrm{d}t$ can be written as 
\begin{equation}\label{eqRandomProcess}
\mathrm{d}N = -(1{-}\pe)\, \mathrm{d}N_\mathrm{c}, 
\end{equation}
where $\mathrm{d}N_\mathrm{c}$ is the number of particles that enter the low-velocity piles in the time interval $\langle t, t{+}\mathrm{d}t\rangle$. In other words, $\mathrm{d}N_\mathrm{c}$ describes the supply of particles into our random process, and, as outlined by Fig.~\ref{figCapture}, it is a function of $\avt$. Under the assumption (indicated by the question mark above the equality sign) that particle settling is a Poisson counting process with intensity $\avt$, we may further write 
\begin{equation}\label{eqPoisson}
\mathrm{d}N = -(1{-}\pe) \,\mathrm{d}N_\mathrm{c} \stackrel{?}{=} - N\, \avt\, \mathrm{d}t .
\end{equation} 

Eq.~\eqref{eqPoisson} can be validated numerically. However, due to intrinsic fluctuations of thermal convection, it is convenient to integrate Eq.~\eqref{eqPoisson} in time and compare the integral quantities instead,
\begin{equation}\label{eqPoisson2}
(1{-}\pe) \stackrel{?}{=} \avt\, \frac{\int_0^t N(t')\,\mathrm{d}t'}{\int_0^t \mathrm{d}N_\mathrm{c}} =: \avt\,\tc.
\end{equation}
The RHS of Eq.~\eqref{eqPoisson2} is plotted in black in Fig.~\ref{figEscape}a. The ratio of the two integrals is labeled as $\tc$ because it is the average time between particle captures, that is, between repeated entries into the low-velocity piles.

By comparing $1{-}\pe$ and $\avt\,\tc$, we observe that for particles with sufficiently small terminal velocities ($\avt/\vm \lesssim 0.02$), the governing equation of the Poisson process is well satisfied (compare blue and black symbols in the inset of Fig.~\ref{figEscape}a). This confirms the idea underlying the derivation in Eq.~\eqref{eqNokes}. Note that for a small value of $\pf$ our description reduces to the one by \citet{Martin1989}: piles with a vanishing $\pf$ factor become only thin layers at the top and bottom boundaries, from which there is no escape (i.e.~$\pe{=}0$). When $\pe=0$, Eq.~\eqref{eqPoisson} reduces to $ N(t)\,\avt\,\mathrm{d}t \stackrel{?}{=} \mathrm{d}N_\mathrm{c}$, where the left hand side expresses the particle flux that can be expected through any horizontal plane if uniformly distributed particles drift vertically with the Stokes velocity. In the limit of $\pf\rightarrow 0$, the right hand side is the supply of particles into the boundary layer. According to the assumptions of \citet{Martin1989}, these two particle fluxes are equal (Eq.~\eqref{eqNokes}). 

Our results show that this assumption is not valid generally: the value of $\avt\,\tc$ reaches up to 4 for light particles in the set C$^{10}_{50}$ (black circles in Fig.~\ref{figEscape}a). Indeed, unlike the probability $1{-}\pe$, the value of $\avt \tc$ is not limited by 1: its value merely evaluates how frequent the transport of particles between fast and slow regions of the flow is. The difference between the ratio $\avt \tc / (1{-}\pe)$ and 1 is then the observed deviation from a Poisson counting process (Fig.~\ref{figEscape}b). 

Already for terminal velocities exceeding $\sim 0.02\,\vm$, the RHS of Eq.~\eqref{eqPoisson2} is larger than the probability $1{-}\pe$. This explains why the exponential law \eqref{eqNokes} fails: it is too difficult for particles to enter the low-velocity piles. In other words, the transport of particles into the sluggish regions, where separation from the fluid flow takes place, is much slower than what one would predict when simply assuming that the particles drift vertically with the speed $\avt$.    

Comparing $\avt\,\tc$ and $1{-}\pe$ provides a unified description of all four regimes. Dust-like: Both quantities are equal and $N=N_0\exp(-\avt\,t)$ describes particle settling well. Transitional: The supply of particles into the piles is too small ($\avt\,\tc > 1{-}\pe$), which implies that settling is slower than that predicted by Eq.~\eqref{eqNokes}, and the disagreement increases as $\avt/\vm$ increases. Bi-linear: The supply of particles is still too small, but increases quickly as $\avt/\vm$ further rises, effectively reducing the difference between $|\vt|\,\tc$ and $1{-}\pe$. This is because the particles are increasingly efficient in penetrating the fluid flow. Stone-like: Eventually, $\avt\,\tc$ becomes smaller than $1{-}\pe$, and the settling curves become faster than $\exp(-\avt\,t)$, soon reaching the Stokes's law instead. These newly defined regime boundaries do not exactly overlap those used in Fig.~\ref{fig4stagioni}, but both definitions are in a rough agreement.

While $\pe$ and $\tc$ depend on the value of $\pf$ that is used in the definition of the low-velocity piles (i.e.~$ |\bm{u}|/\vm < \pf$), their ratio does not (Fig.~\ref{figEscape}b). The quantity $|\vt|\, \tc/(1-\pe)$ can thus be used to construct a general model, extending the exponential law \eqref{eqNokes}. First, we fit the ratio with a skew normal distribution $f$:  
\begin{equation}\label{eqpoly}
\frac{\tc \avt}{1-\pe} = f(|\vt|/\vm) := 0.5 + \frac{A}{\sigma \sqrt{2\pi}}\left[ 1 + \mathrm{erf} \left( \frac{\lambda(x-\mu)}{\sigma\sqrt{2}} \right) \right]\exp{\frac{-(x-\mu)^2}{2\sigma^2}}  \\
\end{equation}
where $x=|\vt|/\vm$ and $\mathrm{erf}$ is the error function. The amplitude $A$ is prescribed as
\begin{equation}
A:= \frac{0.5\, \sigma\sqrt{2\pi}}{\left[ 1 + \mathrm{erf} \left( \frac{-\lambda\mu)}{\sigma\sqrt{2}} \right) \right]} \exp{\frac{\mu^2}{2\sigma^2}},  
\end{equation}
in order to get the observed match between $\avt \tc$ and $(1-\pe)$ in the limit of zero terminal velocity (i.e.~$f=1$ for $\avt/\vm \rightarrow 0$). The remaining parameters, $\mu, \sigma,$ and $\lambda,$ are obtained by fitting the data (see Fig.~\ref{figEscape}b). 

The analytic prescription \eqref{eqpoly} allows us to continue in describing particle settling as a random process. Setting $\mathrm{d}N_\mathrm{c} = N \mathrm{d}t / \tc$ in Eq.~\eqref{eqRandomProcess}, we have
\begin{equation}
-\mathrm{d}N = (1{-}\pe)\, \mathrm{d}N_\mathrm{c} = \frac{(1{-}\pe)}{\tc}\,N \mathrm{d}t
 \quad \Rightarrow \quad \der{N}{t} = \frac{-|\vt|\,N }{f(|\vt|/\vm)} \label{eqOurmodelder}
\end{equation}
Eq.~\eqref{eqOurmodelder} has exponential solution:
\begin{equation}\label{eqOurmodel}
\frac{N(t)}{N_0}= \exp\left(\frac{-|\vt| t}{ f(|\vt|/\vm) }\right).
\end{equation}
Equation \eqref{eqOurmodel} establishes an extension to the solution \eqref{eqNokes}. It is valid also for $\avt/\vm>0.02$, and bridges the gap between existing analytic solutions for small and large terminal velocities (i.e.~the $\avt\rightarrow 0$ and $\avt\rightarrow \infty$ limits). Note that the domain depth $H$ must be added to the denominator when $|\vt| t$ are to be replaced by their dimensional counterparts $|\vt u^*| \tau$.
 
Apart from Eq.~\eqref{eqNokes}, \citet{Martin1989} also develop a more sophisticated theory, in which $c(0)$ is not simply the average concentration $N/A H$. By assuming a depth-dependent concentration $c=c(z)$ whose temporal changes are governed by the diffusion equation \citep[e.g.][]{Bartlett1969}, \citet{Martin1989} find solutions for $c(0)/\bar{c}$ for several flow and particle parameters, with $\bar{c}$ being the average concentration $N/A H$ (see their Table 2). Note that inserting $c(0) \neq \bar{c}$ into Eq.~\eqref{eqNokes} is exactly analogous to our Eq.~\eqref{eqOurmodel}, with $c(0)/\bar{c}$ being analogous to our $1/f$. While \citet{Martin1989} predict $c(0)/\bar{c}>1$ for particles with a non-vanishing Stokes velocity, we obtain the exact opposite, $f>1$. However, the experimental measurements of \citet{Martin1989} are in agreement with our results, as they systematically measure the settling rates to be slower than Eq.~\eqref{eqNokes} predicts, especially for particles with $\vt/\vm\approx 0.5$. \citet{Martin1989} acknowledge the discrepancy between their theoretical prediction and measurements, and speculate that it may be related to the breakdown of the assumption of one-dimensionality. In particular, they anticipate that the large scale circulation in the fluid could be responsible for the failure of the one-dimensional turbulent diffusion theory, which is exactly what we observe.

The misfit between Eq.~\eqref{eqOurmodel} and the observed settling curves never exceeds 30\%, with the largest error occurring for particles with $0.3<\vt/\vm<1.0$. This is not surprising - already from Fig.~\ref{fig4stagioni}b it is clear that the settling curves are not exponential when $\vt/\vm\gtrsim 0.3$. Nevertheless, Eq.~\eqref{eqOurmodelder} is still useful for estimating the characteristic time of complete sedimentation for all particle types (see Fig.~\ref{figFinal} at the end of this section and the accompanying discussion).


The imperfect fit of the observed settling curves is caused by the fact that $\tc$ is not a function of $\avt$ only, but it is also a function of time, $\tc=\tc(\avt,t)$. For instance, in the stone-like regime, the number of particles $N$ follows $N(t) \approx N_0 \times (1-\vt\,t)$, and $\mathrm{d}N_\mathrm{c} \approx N_0\,\vt \mathrm{d}t$. Eq.~\eqref{eqPoisson2} then yields $\tc \approx \max (1/\vt{-}t/2,\, 0.5/\vt)$, with $0.5/\vt$ being the end value that is reached after all particles have settled (see the black line in Fig.~\ref{figTavg}, resp.~the dotted line in Fig.~\ref{figEscape}b). Eq.~\eqref{eqOurmodel}, on the other hand, assumes that the average time between particle captures, $\tc$, does not vary in time (see the definition of $\tc$ in Eq.~\eqref{eqPoisson2} versus its use in Eq.~\eqref{eqOurmodelder}).

In Fig.~\ref{figTavg} we show the temporal evolution of $\tc$ for selected $\vt/\vm$ ratios. Regardless of the value of $\vt/\vm$, $\tc$ equilibrates before $t \approx 100$ (note that the x-axis is logarithmic -- for most of the depicted time-window $\tc$ is steady). By $t\approx 100$, however, most of the particles in the stone-like and bi-linear regimes have already settled (dashed lines). Refining Eq.~\eqref{eqOurmodel} would thus require accounting for the time dependence of $\tc$. 

With the black line in Fig.~\ref{figTavg} we plot $\max (1/0.3{-}t/2,\, 0.5/0.3)$, which is the theoretical value of $\tc$ obtained for Stokes's settling of particles with $\vt=0.3$ (see above). The initial, short-lived drop of the actual evolution of $\tc(t)$ (dark blue line) is caused by accelerating to $\vt$ from zero (the average vertical velocity of the fluid, $\int \bm{u}_z \mathrm{d}V$, and thus also the initial average velocity of the particles, is zero). In the transitional and dust-like regimes, on the other hand, $\tc(t)$ has only a mild temporal evolution and the transient phase is also less relevant, because the particles take longer to settle. Most of the particles with small terminal velocities settle when $\tc(t)$ is completely steady, allowing Eq.~\eqref{eqOurmodel} to provide a good fit to the observed settling curves ($<10$ \% error for $\vt/\vm<0.1$).

\begin{figure}[ht]
\includegraphics[scale=0.33]{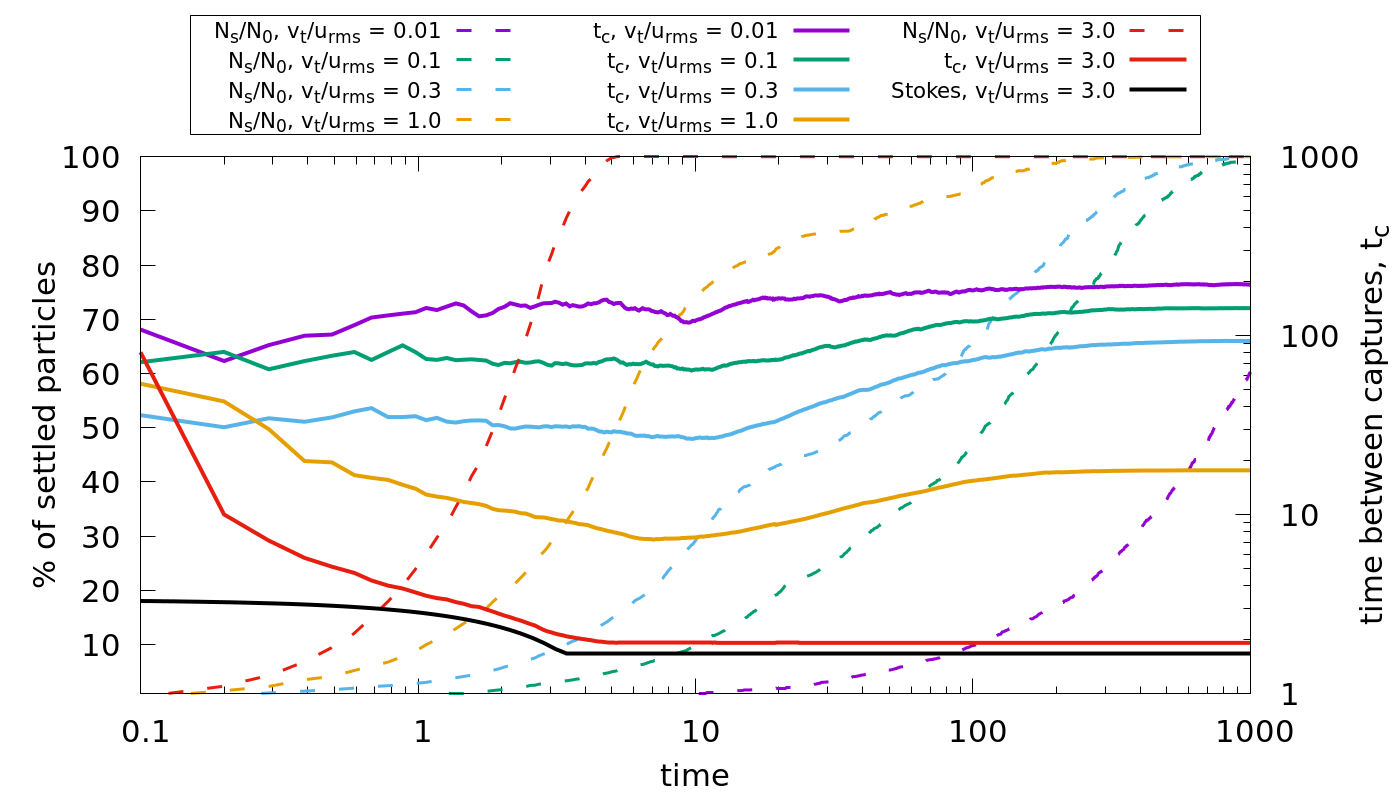}\caption{Solid lines show the temporal evolution of the average time between captures, $\tc$, for selected $\vt/\vm$ ratios. For a comparison, the characteristic time of a flow overturn is 5. The black line is equal to $\max (1/0.3{-}t/2,\, 0.5/0.3)$. Dashed lines show the corresponding percentages of settled particles.}
\label{figTavg}
\end{figure}

In Fig.~\ref{figFinal} the above results are summarized. In the literature, the terminal velocity $\vt$ is used as a measure of the settling rate of particles in a convective flow. This implies that the time required for a complete sedimentation of all particles is $1/(\vt)$, or $H/(\vt u^*)$ in terms of dimensional quantities. Fig.~\ref{figFinal} shows the factor $F$ by which $1/(\vt)$ has to be multiplied in order to obtain the correct settling time. Since Eq.~\eqref{eqOurmodel} is an exponential law, the predicted time of complete sedimentation is, in principle, infinite. To circumvent such inconvenience, we compute the time until 95\% of particles have settled and divide the resulting value by $0.95/\vt$ (i.e.~normalize by the respective terminal time). The plotted factor, $F$, is therefore equal to:
\begin{equation}\label{eqfactor}
F:= \frac{-\log (0.05)}{0.95} f(\avt/\vm).
\end{equation}

\begin{figure}[ht]
\includegraphics[scale=0.65]{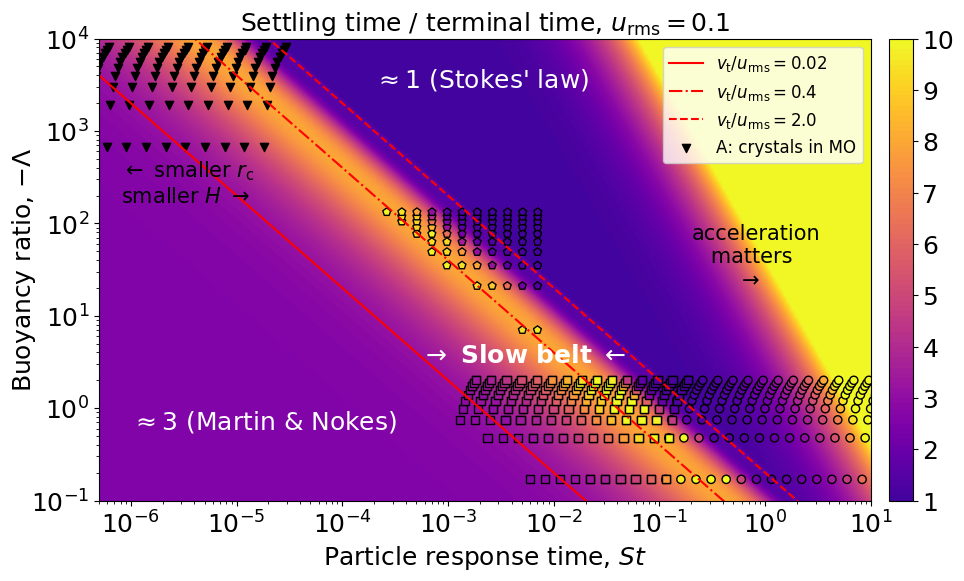} \caption{Normalized settling time $F$ as a function of buoyancy ratio $\Lambda$ and particle response time $St$. The settling time is computed with our analytic model \eqref{eqOurmodel}, while squares, circles, and pentagons are colored according to the actual settling times obtained from the numerical simulations C$^{10}_{50}$, xC$^{10}_{50}$, and B$^{10}_{50}$ respectively. The region of relatively slow settling (``slow belt'' in the figure) occupies the area with $0.02 \lesssim \avt/\vm \lesssim 2.0$. Black triangles correspond to the parameters from Table \ref{magpar}. The colour scale is clipped in order not to be dominated by the top right region, where settling occurs as if there was no convection.}
\label{figFinal}
\end{figure}

The settling time $/$ terminal time ratio, $F$, is computed with the use of Eq.~\eqref{eqOurmodel} when $\vt/\vm<2$ and from Eq.~\eqref{eqrestint} when $\vt/\vm>2$. We restrict the applicability of our analytic model, because for $\vt/\vm>2$ the Stokes' formula \eqref{eqrest} is more accurate and physically appropriate. The parameters $\mu,\sigma,$ and $\lambda$ used in Fig.~\ref{figFinal} come from the set C$^{10}_{50}$ (see Table \ref{tableai} for the respective values, Fig.~\ref{figFinal} is plotted for heavy particles only).

Solid and dashed red lines in Fig.~\ref{figFinal} are isolines of $\vt$ that correspond to the regime boundaries. Since $\vm=0.1$, the respective values are $\vt=0.002$, 0.04, and 0.2. For a more vigorous flow these values must be adjusted accordingly (see Section \ref{sec:MO}).

On top of the analytic prediction \eqref{eqOurmodel} we plot the settling times observed in the simulation sets B$^{10}_{50}$, C$^{10}_{50}$, and xC$^{10}_{50}$ (pentagons, squares, and circles respectively). Only the particle types for which the simulations have reached at least 95\% settling are plotted. The difference between Eq.~\eqref{eqOurmodel} and the numerical simulations is typically less than 3\%. One exception is the vicinity of $\avt/\vm\approx 0.6$, where the settling rates are very small during the second-stage of the bi-linear regime, which significantly prolongs the settling time with respect to expectations (up to 40\% discrepancy in the value of $F$). Nevertheless, $N_\mathrm{s}(t_{95})/N_0 - 0.95$ stays below 7\% for all particle types, ensuring reasonable accuracy of Eq.~\eqref{eqOurmodel} for general applications. Here, $t_{95}$ is the settling time computed by setting $(N_0-N)/N_0 = 0.95$ in Eq.~\eqref{eqOurmodel}.

The region labeled ``slow belt'' is characterized by increased settling times, with the average settling rates being up to 13 times slower than Stokes's law predicts in the present conditions $Ra=10^{10}$ and $Pr=50$. With the exception of the above mentioned discrepancy between the theoretical prediction and measurements of \citet{Martin1989}, its existence is not reported in previous literature, which calls for experimental and 3D numerical confirmation of our findings.

\section{Results: importance of the background flow}\label{sec:background}

In the dust-like and stone-like regimes, the settling curves are robust with respect to properties of the background flow. In the transitional and bi-linear regimes, on the other hand, the above results suggest that settling curves depend on large-scale circulation of the fluid (Fig.~\ref{figBilinear}). Thermal convection of an isoviscous fluid naturally leads to the formation of low-velocity piles similar to those depicted in Fig.~\ref{figEyes} (see also Discussion), but their coherence and erosion depend on the values of $Ra$ and $Pr$. In this section we analyze the interplay between particle settling and the velocity structure of thermal convection.

\subsection{Horizontal distribution of settled particles}\label{sec:horizontal}
The escape probability $\pe$ is close to zero already for $\avt/\vm\gtrsim 0.3$ (Fig.~\ref{figEscape}), and the settling problem is thus mostly reduced to measuring the transport of particles from the bulk of convection into the low-velocity piles. Here we show that the near-boundary regions depicted in green in Fig.~\ref{figEyes}, i.e.~the piles with $\pf=0.5$, act as dominant sinks for most particle types. 

In Section \ref{sec:Bilinear} we demonstrated that in the second stage of the bi-linear regime, heavy particles hover inside a cluster of upwellings (Fig.~\ref{figBilinear}). Eventually, these particles settle in the underlying low-velocity pile, whose edges are a continuous source of the plumes that keep lifting the particles. Note that settling below upwellings is somewhat counter-intuitive -- one could naturally expect heavy particles to concentrate below major downwellings. 

\begin{figure}[ht]
\includegraphics[scale=0.33]{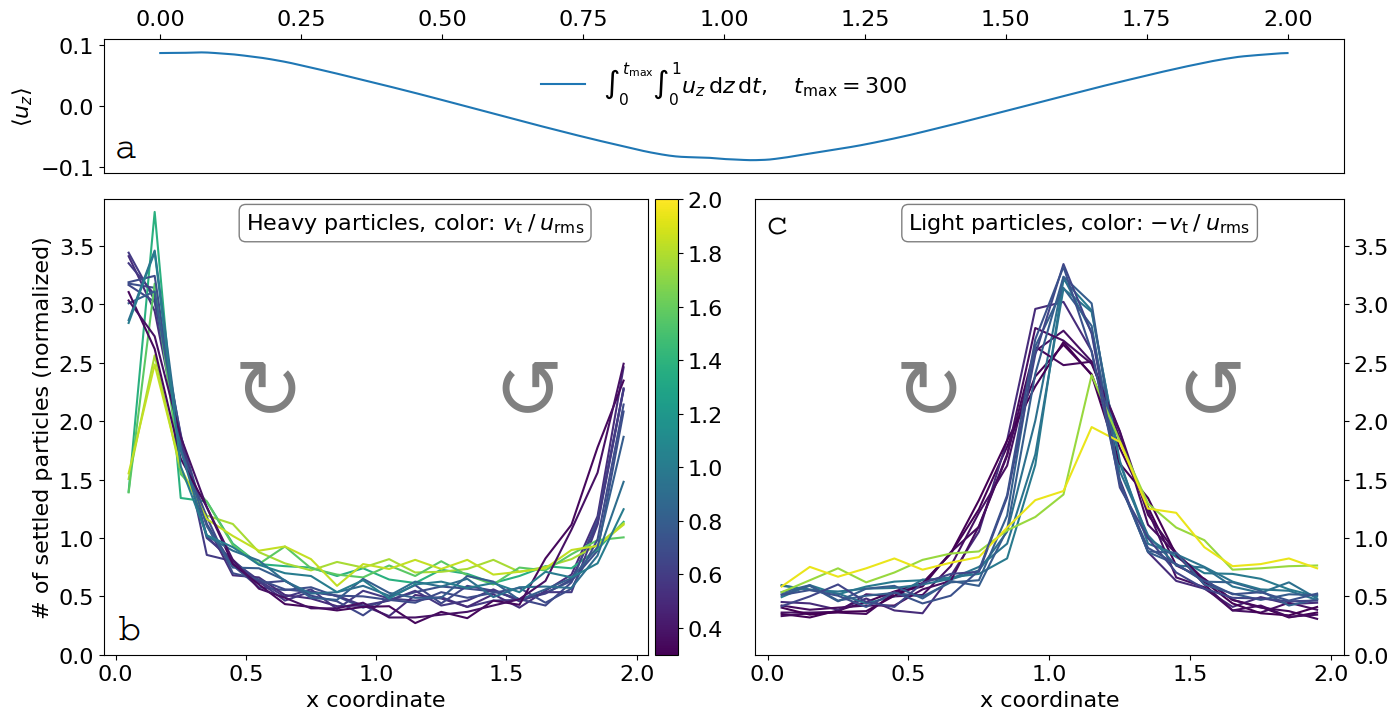}
\caption{Horizontal distribution of settled particles in the simulation set C$^{10}_{50}$. The $\vt/\vm$ ratio is depicted in color, and only the bi-linear range, $0.3<\avt/\vm<2.0$, is selected. (a) Vertically averaged vertical component of the fluid velocity, $u_z$. The quantity is further averaged over time, with the time window of integration covering complete sedimentation of the respective particles. (b) Heavy particles. (c) Light particles. Gray symbols indicate the flow direction of two major convection cells. The settled particles are sorted into 20 equally wide bins and the plotted value is normalized by the initial number of particles, i.e.~divided by $N_0 / 20$. }
\label{figLatBl}
\end{figure}

Fig.~\ref{figLatBl} shows the horizontal distribution of settled particles. The distribution combines information from both stages of settling, but we note that the non-uniformity is produced in the second stage only. In the first stage of the bi-linear regime, the relatively fast sedimentation along with the horizontal drag associated with large-scale convection rolls ensure that particles are distributed evenly across the bottom and top boundaries (see also Video S1 in Supplementary material).

For brevity, and because the horizontal distribution of light particles is analogous to the distribution of heavy particles, we will discuss only the heavy particles in this section. Top panel of Fig \ref{figLatBl} shows the vertically- and time-averaged vertical component of the velocity field. Peaks of the function correspond to the edges of large-scale convection cells, whose centres are indicated by the grey symbols in panels b) and c).

\begin{figure}[ht]
\includegraphics[scale=0.33]{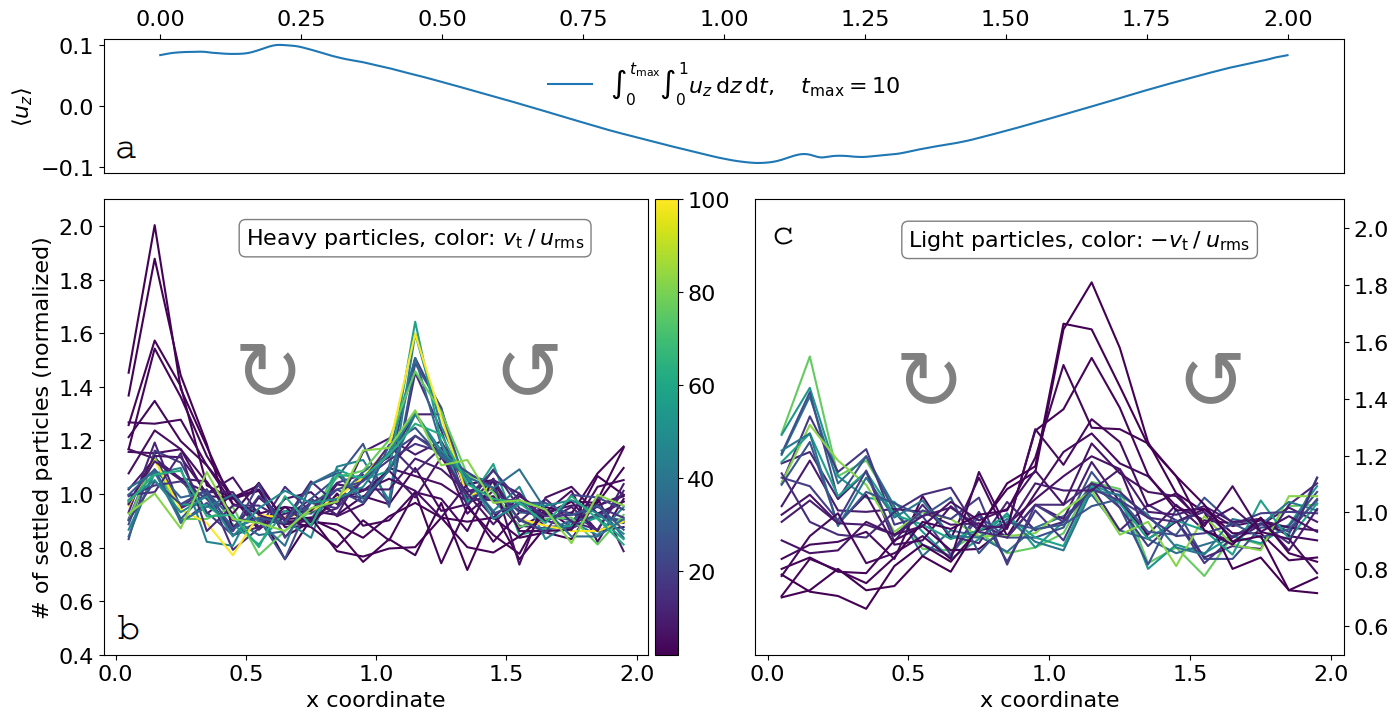}
\caption{As Fig.~\ref{figLatBl}, only here we show the horizontal distribution of settled particles in the simulation set xC$^{10}_{50}$. }
\label{figLatStone}
\end{figure}

Fig.~\ref{figLatStone} shows that also stone-like particles settle preferentially in the large low-velocity pile located at the edges of the model domain (as long as the $\vt/\vm$ ratio is $\lesssim 10$ -- see the dark blue lines). Due to their large sinking velocity, stone-like particles efficiently cut through the bursts of upwellings. Yet, the large low-velocity pile still acts as a sink. This is because, due to the large-scale rolls, the fluid velocities in the lower-half of the model domain generally point towards the side edges, which sways the sinking particles into that direction. 

For the particle trajectories to be close to ``ballistic'', i.e.~to follow the analytic solution \eqref{eqrest} exactly, the $\avt/\vm$ ratios of much larger than 2 are necessary. Even when Eq.~\eqref{eqrest} effectively governs particle dynamics, the horizontal distribution of settled particles is not uniform. The fluid flow then enters Eq.~\eqref{eqrest} through the initial velocities, $\tilde{\bm{v}}(\bm{x},t{=}0) = \bm{u}(\bm{x},t{=}0)$. Because of large-scale circulation, ballistic particles in the upper part of model domain are generally injected in the direction of major downwellings and have sufficient time to move laterally. Particles injected into the lower half of the domain, on the other hand, do not have the time to move to the sides. As a result, the horizontal distribution reverses for $\avt/\vm\gtrsim 30$, with most of the heavy particles settling below the major downwelling (light green and yellow curves in Fig.~\ref{figLatStone}).

For particles with $0.02<\avt/\vm<0.3$, we plot the horizontal distribution of settling in Fig.~\ref{figLatTr}. In the transitional regime, particles still ``see'' the low velocity piles, but the settling events become horizontally uniform as $\avt\rightarrow 0.02$. The particles experience on average over twenty flow overturns between the repeated crossings of the low-velocity piles (see Fig.~\ref{figTavg}). Typically, they circulate in large convection cells, waiting to enter the slow regions through small-scale irregularities of the flow that are produced by births of new plumes (see also the concentration of grey and black dots inside the green regions in Fig.~\ref{figEyes}). As discussed in Section \ref{sec:Ra}, for $Ra=10^{10}$ the low-velocity piles are particularly coherent and do not move horizontally (see also the top panels of the figures in this section), which makes it difficult to enter them. Under such conditions, the particle supply into the piles can be particularly small and the horizontal distribution of settling events can be highly non-uniform.

\begin{figure}[ht]
\includegraphics[scale=0.33]{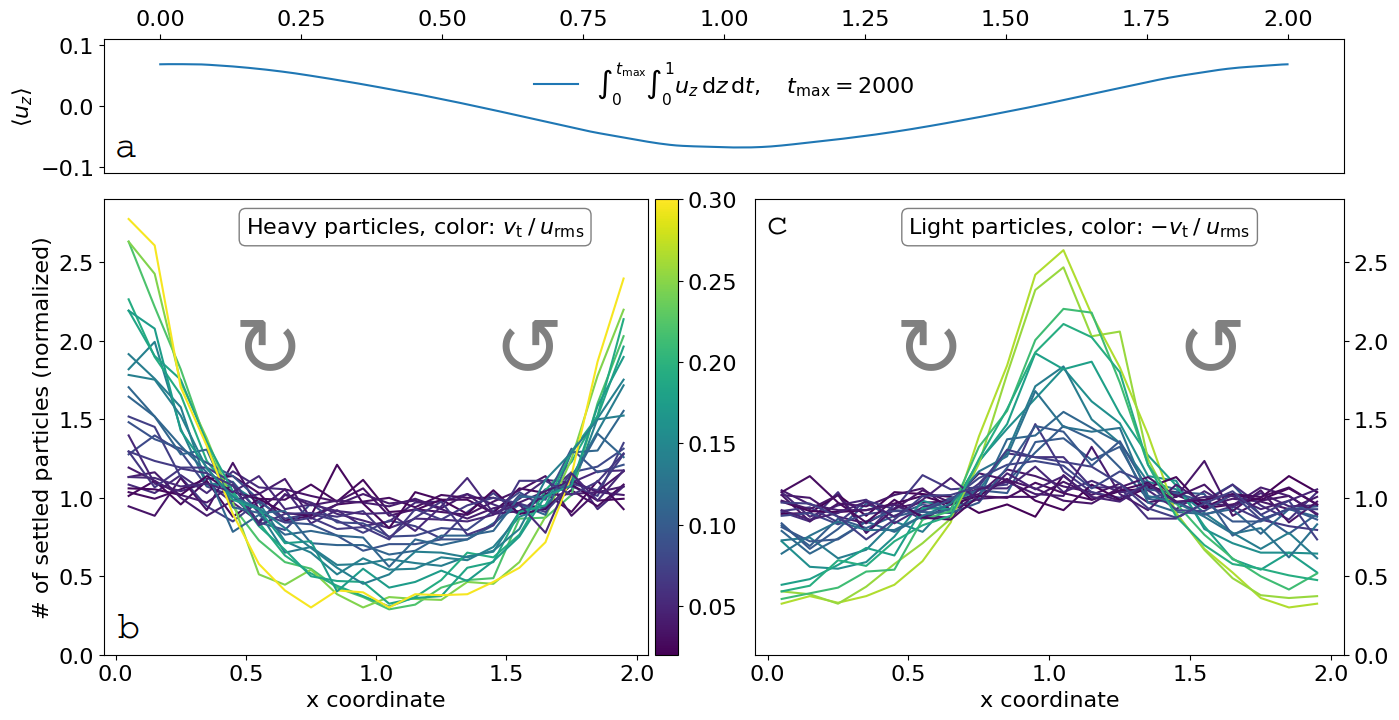}
\caption{As Fig.~\ref{figLatStone}, only here we analyze the transitional regime, $0.02<\avt/\vm<0.03$.}
\label{figLatTr}
\end{figure}

In the dust-like regime, the horizontal distribution of settled particles is uniform. These, almost tracer-like particles remain suspended in the fluid for very long times, repeatedly entering and leaving the low-velocity piles (Fig.~\ref{figCapture}). The escape probability $\pe$ approaches 1 as $\avt/\vm\rightarrow 0$, which means that for particles with small terminal velocities the piles become transparent. Convective motions inside the piles, though relatively slow, are still fast enough to drag along the particles with a vanishing response time $St$ (i.e.~the dust-like particles). The only structure where such particles can separate from the fluid becomes the thin, laterally uniform part of the no-slip boundary layer.

\subsection{Motion of light particles towards vortices ($\beta$-effect)}\label{sec:beta}

Describing the settling behaviour with the help of $\avt/\vm$ only is a crucial reduction of the five-dimensional model parameter space. Given the complexity of the problem at hand, such description can only be used as a first-order approximation.

One difficulty appears already in Fig.~\ref{figEscape}a, since there is clearly an asymmetry between light and heavy particles that have the same amplitude of the terminal velocity. For $Ra=10^{10}$, the differences are small and may partially result from a random asymmetry of the up-and downwelling regions. For $Ra=10^{12}$, the asymmetry becomes a prominent feature (Fig.~\ref{figBlBeta}). Its underlying mechanism is explained below.

\begin{figure}[ht]
\includegraphics[scale=0.165]{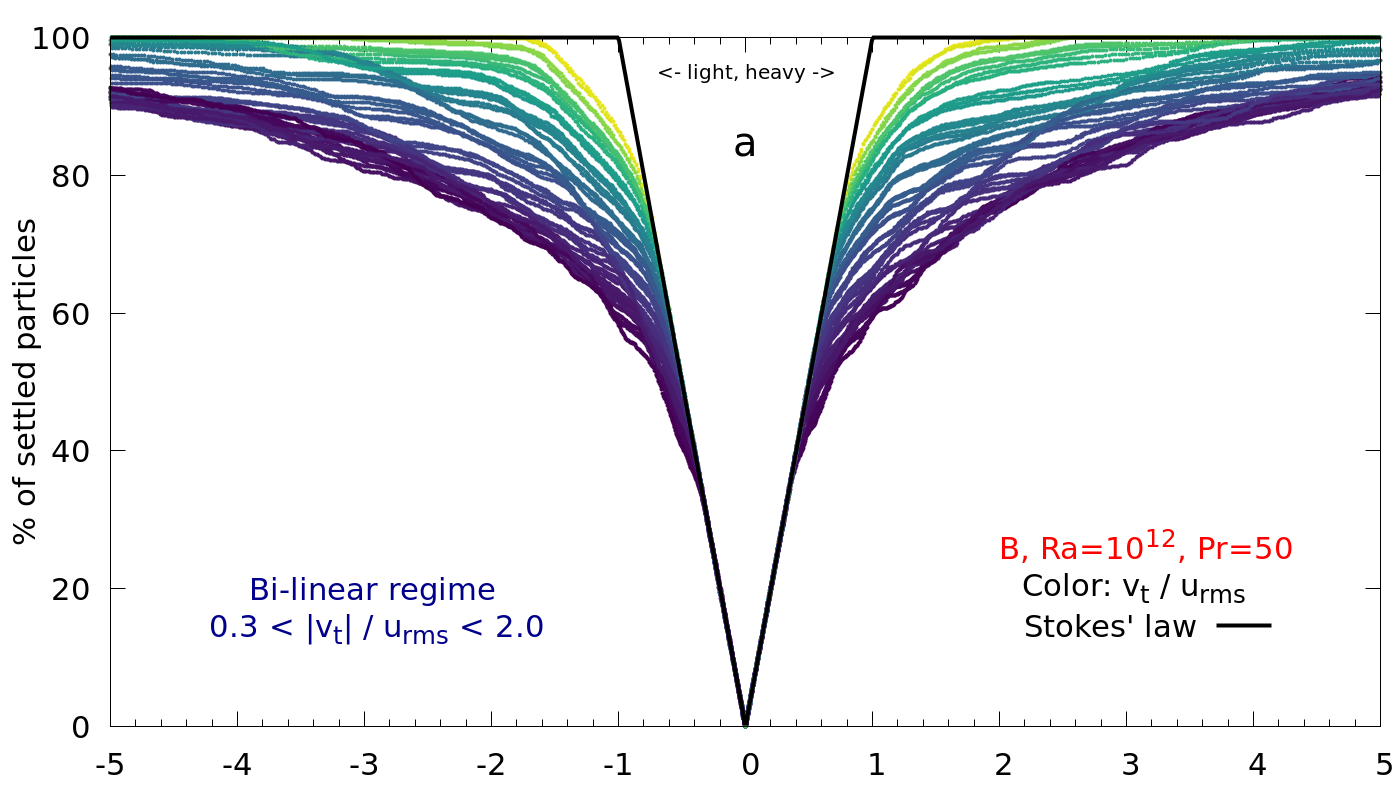}
\includegraphics[scale=0.165]{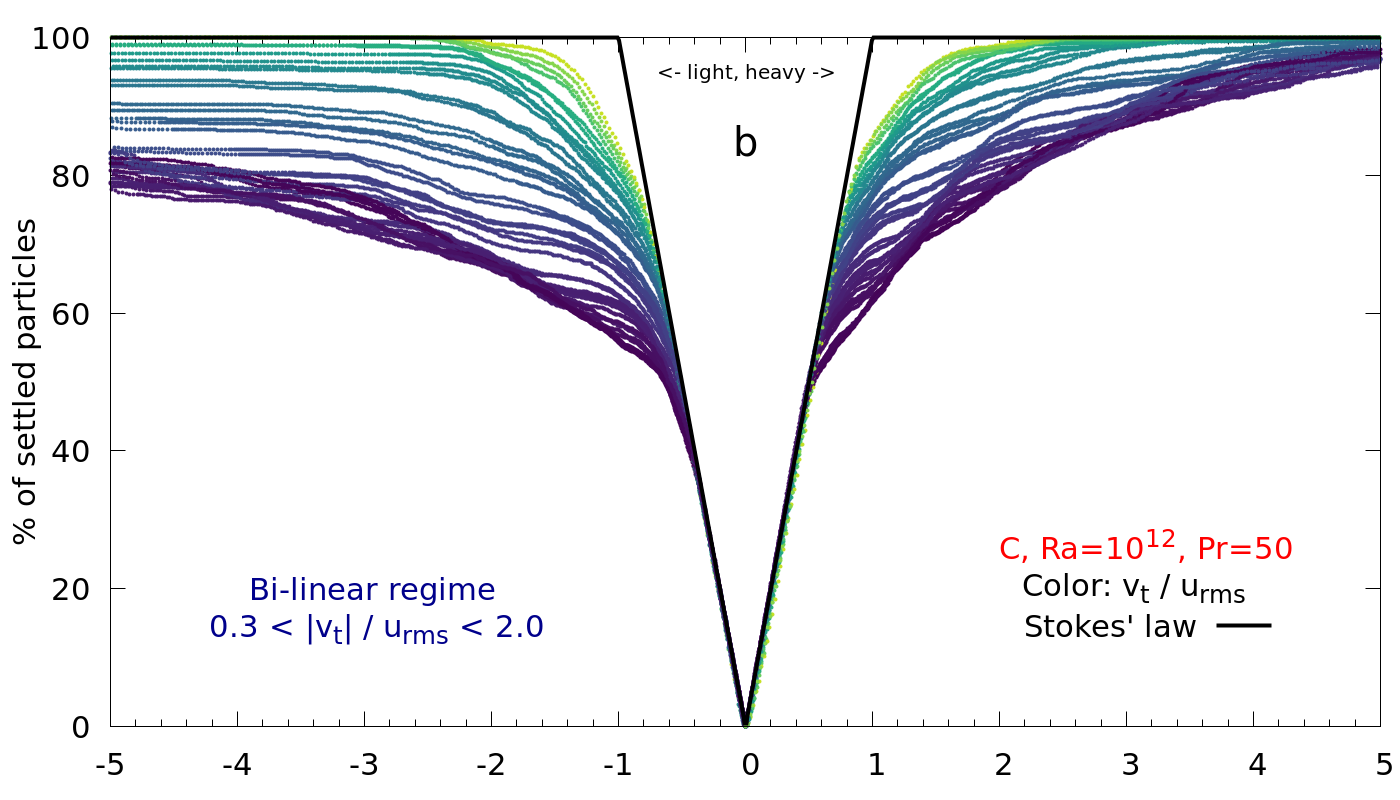}
\caption{Same as Fig.~\ref{fig4stagioni}b, only here for the simulation sets B$^{12}_{50}$ (left) and C$^{12}_{50}$ (right). While B$^{12}_{50}$ contains particle types with the modified density ratio $\beta$ very close to 1 and the resulting settling curves show a strong light/heavy symmetry, for the simulation C$^{12}_{50}$ the situation is different. Due to the $\beta$-effect, heavy particles are pushed away from flow vortices and light particles towards them. This results in faster settling for heavy particles and an extreme flattening of the settling curves of light particles.}
\label{figBlBeta}
\end{figure}

Due to the term $\beta {D\mathbf{u}}/{Dt}$ in Eq.~\eqref{eqnonDMR}, heavy particles ($\beta{<}1$) have a tendency to move away from strong flow vortices \citep{Eaton1994}. Indeed, in Fig.~\ref{figHighRa}, where we show a snapshot from the simulation C$^{12}_{50}$, there is a reduced concentration of heavy particles at the edges of the particle cloud, i.e.~close to the centres of the two largest convection rolls (in Fig.~\ref{figBilinear} this effect is not observed because the Reynolds number is too small).

\begin{figure}[ht]
\includegraphics[scale=0.55]{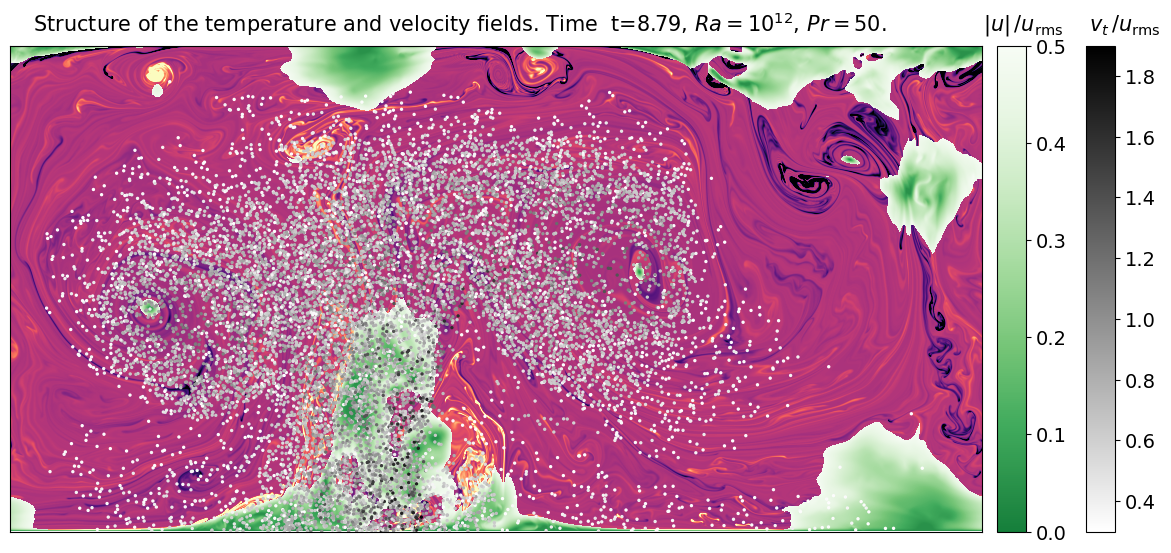}
\caption{Similar to Figs.~\ref{figBilinear} and \ref{figEyes}, only here for the simulation set with the highest Rayleigh number: C$^{12}_{50}$. The temperature scale is clipped, with the respective range being depicted in Fig.~\ref{figBeta} below. The snapshot is taken at $t=0.4/(0.3\,\vm)$, i.e.~at the time for which the terminal distance is 0.4 for the settling curve with $\vt/\vm=0.3$ (as in Fig.~\ref{figBilinear}). See Supplementary material, Video S3.}
\label{figHighRa}
\end{figure}

Light particles, on the other hand, move toward flow vortices thanks to the $\beta {D\mathbf{u}}/{Dt}$ term \citep{Maxey1987}. In Fig.~\ref{figBeta}, we show only the particles with $\vt/\vm\approx -0.3$ (the light/heavy asymmetry seems largest for this value, see Fig.~\ref{figBlBeta}). Around strong and long-lived vortices, there is an increased concentration of light particles. These particles are trapped until the respective vortices vanish, which explains the enhanced flattening of the settling curves. 

Due to the way particle sets B and C are constructed, the range of $\beta$ is approximately $\langle 0.99,1.01 \rangle$ for one and $\langle 0.6,3.0 \rangle$ for the other. Since the outward (resp.~inward) motion of heavy (resp.~light) particles depends on how much $\beta$ departs from unity, the effects encountered in Figs \ref{figHighRa} and \ref{figBeta} for the set C$^{12}_{50}$ do not occur in B$^{12}_{50}$. In terms of the settling curves, B$^{12}_{50}$ shows a symmetry between light and heavy particles, while C$^{12}_{50}$ exhibits a slightly enhanced settling of heavy particles and a significantly delayed settling of light particles (Fig.~\ref{figBlBeta}).  

\begin{figure}[ht]
\includegraphics[scale=0.55]{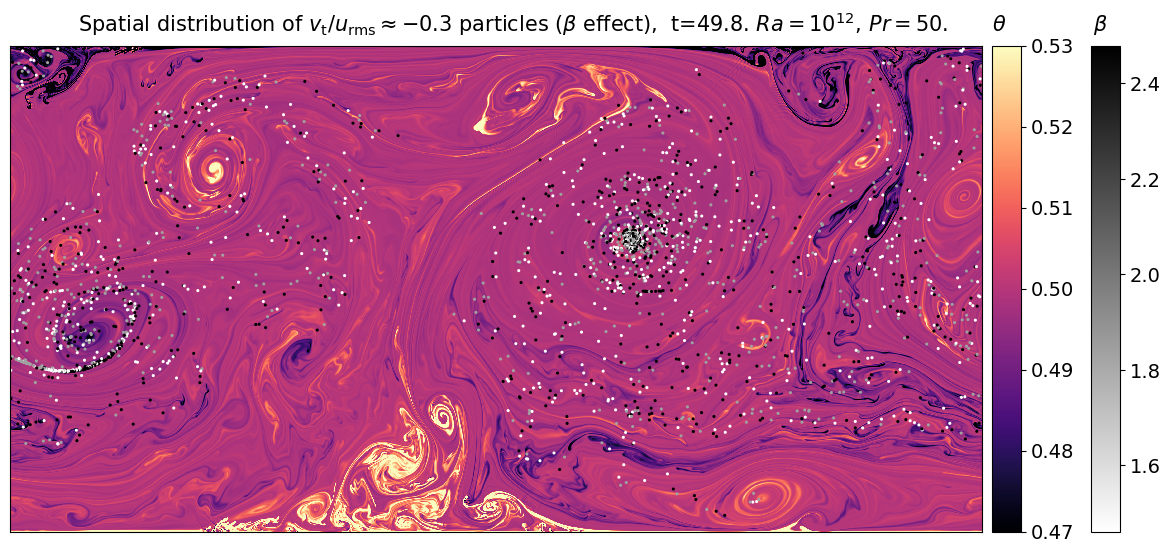}
\caption{Demonstration of the $\beta$-effect. When convective vigor is high, particles from the transitional and bi-linear range, $0.02<\avt/\vm<2.0$, have a tendency to move away, resp.~towards strong flow vortices. Here we show the increased concentration of light particles with $\vt/\vm = -0.3 \pm 0.02$ near long-lived vortices that have developed in the simulation set C$^{12}_{50}$. See Supplementary material, Video S4.}
\label{figBeta}
\end{figure}

Each simulation set contains several particle types with roughly the same $\avt/\vm$ ratio. For C$^{12}_{50}$ and $\vt/\vm = -0.3 \pm 0.02$, the modified density ratio $\beta$ ranges from 1.4 to 2.5 (see Fig.~\ref{figStLspace}). The respective particle types are shown in Fig.~\ref{figBeta}. For these few particle types, the concentration in the vicinity of stable vortices seems similar, and also their settling curves are comparable. Based on this particular example, we can crudely conclude that for $\beta\in \langle 1.4,2.5 \rangle$ the $\beta$-effect is similarly strong and for $\beta<1.01$ no effect is observed. A more detailed analysis should be the subject of future work.

\begin{figure}[ht]
\includegraphics[scale=0.49]{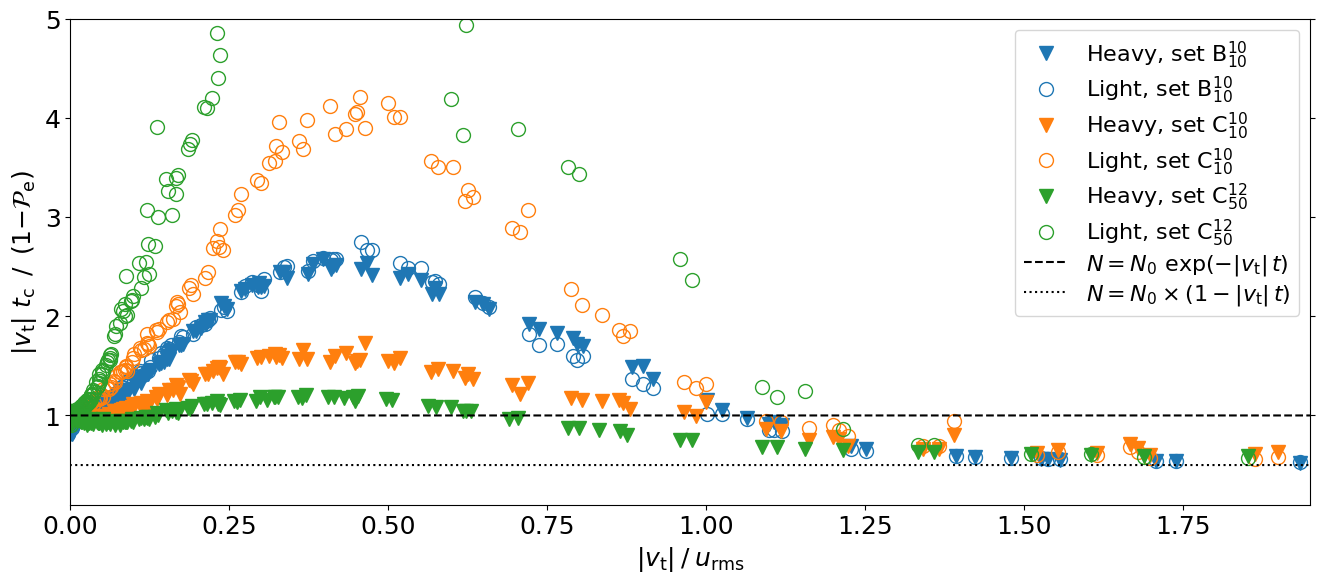}\caption{ $\avt \tc / (1{-}\pe)$ ratio for the simulation sets B$^{10}_{10}$, C$^{10}_{10}$, and C$^{12}_{50}$. Light particles (circles) experience slower settling than the heavy ones, as long as the $\beta$-effect comes into play (sets labeled as C). The separation between the ratios for light and heavy particles increases with the Rayleigh number (cf.~the orange and green symbols).}
\label{figRatio}
\end{figure}

One way to quantify the $\beta$-effect as a function of $\vt/\vm$ is through the $\vt \tc / (1{-}\pe)$ ratio. Already in Fig.~\ref{figEscape} there was a difference between light and heavy particles, and the gap further increases as the Reynolds number increases. In Fig.~\ref{figRatio} we plot $\vt \tc / (1{-}\pe)$ for the sets B$^{10}_{10}$, C$^{10}_{10}$, and C$^{12}_{50}$. Consistently with the analysis above, light particles settle very slowly in C$^{12}_{50}$, with $\vt \tc / (1{-}\pe)$ going up to 14 for particle types that are trapped inside vortices for a particularly long time, while there is little difference between the settling rates of light and heavy particles in the set B$^{10}_{10}$. 

Note that for heavy particles the value of $\vt \tc / (1{-}\pe)$ may drop below 1 due to the $\beta$-effect (green triangles in Fig.~\ref{figRatio}). This result can be interesting in view of the debate between \citet{Wang1993} and \citet{Mei1994}: the former observed faster than Stokes' settling of heavy particles due to preferential sweeping in downward moving fluid, while the latter did not observe the effect \citep[see also][]{Lavorel2009, Bosse2006}. Here we find an increased settling rate that is higher than predicted by the exponential law of \citet{Martin1989}, but it never exceeds the Stokes' velocity on average. Note, however, that our $Pr$ is always relatively high, limiting turbulence effects.

\subsection{Effects of convective vigor and fluid inertia}\label{sec:Ra}

In this section we analyze how the above results are affected by $Ra$ and $Pr$ of the background flow. In Table \ref{tableai}a we show the parameters $\mu,\sigma,$ and $\lambda$ of the skew normal distribution $f$ for all the simulation sets labeled as B, along with the maximum value of $f$, denoted as $m$, and the value of $\avt/\vm$ at which the maximum is reached. The maximum $m$ can be used as an estimate for how much slower the particle settling can be when compared to the exponential decay $N=N_0\exp(-\avt t)$.

\begin{table}[ht]
\caption{a) Parameters of distribution $f$, no $\beta$-effect.}\label{tableai}
\centering
\begin{tabular}{l c c c c c c c}
\hline
Set & Reynolds n. & $\mu$ & $\sigma$ & $\lambda$ & max.~$m$ & at $\avt/\vm$ \\
\hline
B$^{08}_{50}$ & 100 & 0.04 & 0.64 & 3.46 & 1.5 & 0.32 \\
B$^{08}_{10}$ & 430 & 0.18 & 0.51 & 2.25 & 2.4 & 0.45 \\
B$^{10}_{50}$ & 1440 & 0.18 & 0.43 & 2.43 & 3.2 & 0.39 \\
B$^{10}_{10}$ & 6770 & 0.21 & 0.44 & 1.84 & 2.6 & 0.45 \\
B$^{12}_{50}$ & 21790 & 0.23 & 0.40 & 0.81 & 1.6 & 0.42 \\
\end{tabular} \\
{\small b) Parameters of distribution $f$, with $\beta$-effect.} \\
\begin{tabular}{l c c c c c c}
\hline
Simulation set & $\mu$ & $\sigma$ & $\lambda$ & max.~$m$ & at $\avt/\vm$ \\
\hline
Heavy, C$^{08}_{50}$ & 0.02 & 0.82 & 7.44 & 1.6 & 0.25 \\
Light, C$^{08}_{50}$ & 0.04 & 0.69 & 4.73 & 1.7 & 0.31 \\
Heavy, C$^{10}_{10}$ & 0.11 & 0.60 & 1.93 & 1.6 & 0.44 \\
Light, C$^{10}_{10}$ & 0.25 & 0.39 & 1.77 & 4.0 & 0.46 \\
Heavy, C$^{10}_{50}$ & 0.14 & 0.60 & 4.10 & 3.3 & 0.39 \\
Light, C$^{10}_{50}$ & 0.20 & 0.51 & 3.09 & 4.3 & 0.44 \\
Heavy, C$^{10}_{10}$ & 0.11 & 0.60 & 1.94 & 1.6 & 0.44 \\
Light, C$^{10}_{10}$ & 0.25 & 0.39 & 1.77 & 4.0 & 0.46 \\
Heavy, C$^{12}_{50}$ & 0.87 & 1.01 & -3.50 & 1.1 & 0.42 \\
Light, C$^{12}_{50}$ & 0.28 & 0.35 & 2.01 & 9.6 & 0.47 \\
\end{tabular}
\end{table}

The most important outcome of the comparison in Table \ref{tableai}a is that the critical ratios $\avt/\vm$ that mark the regime boundaries are largely independent of $Ra$ and $Pr$: the maximum of the function $f$, i.e.~the boundary between the transitional and bi-linear regimes, always lies at $\avt/\vm \approx 0.4 \pm 0.1$. With a similar accuracy, the stone-like regime is always obtained for $\avt/\vm \approx 2.0$ (recall that Stokes' settling satisfies $\vt\tc/(1{-}\pe) = 0.5$, see the dotted line in Fig.~\ref{figEscape}b). 

The settling behaviour in the limits $\avt/\vm\rightarrow \infty$ and $\avt/\vm\rightarrow 0$ can be derived analytically regardless of the values of $Ra$ and $Pr$: in the first case, the flow is irrelevant. In the second case, the only requirements are those discussed above in relation to Eq.~\eqref{eqNokes}. The critical values of $\avt/\vm$ for which the settling curves start to substantially deviate from these limits, could, however, strongly depend on the Rayleigh and Prandtl numbers. The fact that there seems to be no such dependence makes the possibility to extrapolate our results to arbitrary thermal flows promising.

The only notable difference between the various simulation sets is the width of the bi-linear and transitional bands, i.e.~the spread of the settling curves in the transitional and bi-linear regimes. The width of the band is directly linked with the maximum value $m$ (compare e.g.~the sets B$^{12}_{50}$ and C$^{12}_{50}$ in Figs \ref{figBlBeta} and \ref{figRatio}), and $m$ depends non-trivially on $Ra$ and $Pr$. Generally, there is a trend between $m$ and the Reynolds number, with $m$ being the largest for $Re \in [10^3,10^4]$.

Increasing the Reynolds number $Re$ enhances the short-wavelength content of the velocity field, and alters the stability of large-wavelength structures. In particular, the low-velocity regions become less stable (see Fig.~\ref{figHighRa}). High convective vigor is capable of tearing the sluggish, boundary-based piles into chunks that are advected into the rest of the fluid and mixed. This results in spatial variations of large-scale circulation and thus in faster settling because the sinking particles spread across a broader area (see Videos S2 and S4 in Supplementary material).

Surprisingly, particles with $0.02<\avt/\vm<2.0$ show faster settling also when $Ra$, resp.~$Re$ decreases. As expected, in the simulation C$^{8}_{50}$ the plumes are thicker and live longer than those in the simulations C$^{10}_{50}$ and C$^{12}_{50}$. The thicker and well separated plumes, however, allow particles to sink in between them, which results in an increased settling rate in the second stage of the bi-linear regime when compared to the reference set C$^{10}_{50}$ (see also Video S5 in Supplementary material).

For sets labeled as C and $Re\geq 10^{2}$, the $\beta$-effect splits the $\vt \tc / (1{-}\pe)$ ratio into two clearly distinct functions. In Table \ref{tableai}b we provide the corresponding sets of parameters $\mu,\sigma,$ and $\lambda$. The difference between $m$ for the heavy and light particles increases with the Reynolds number.

\section{Application to crystallizing magma}\label{sec:MO}

The Rayleigh number of a magma chamber is of the order of $10^9$--$10^{17}$ \citep[e.g.][]{Clark1987}, making our results directly applicable to relatively small volumes of magma. For a global magma ocean, however, $Ra \sim 10^{27}$ (Table \ref{magpar}). First of all, in order to apply our results to this  system, it is necessary to estimate $\vm$ for such an extreme flow regime.

In terms of non-dimensional control parameters, the $\vt/\vm$ ratio can be expressed as:
\begin{equation} \label{eqvtvm}
\frac{\vt}{\vm}=\frac{-\St\,\Lambda\, u^*}{U_\mathrm{rms}}=\frac{-\St\,\Lambda\, \sqrt{\alpha g \Delta T H} \,H}{Re \,\nu} = \frac{-\St\, \Lambda \sqrt{Ra}}{Re\,\sqrt{Pr}},
\end{equation}
where the Reynolds number is defined through the volume-averaged root mean square velocity, $Re:=U_\mathrm{rms} H / \nu$. 

Upon employing a $Re=Re(Ra,Pr)$ scaling, Eq.~\eqref{eqvtvm} can be used to compute the $\avt/\vm$ ratio for various Rayleigh and Prandtl numbers. Relationships between Reynolds number and Rayleigh and Prandtl numbers have been obtained based on various experimental, numerical, and theoretical work. Here we adopt the Grossmann-Lohse theory \citep{Grossmann2000}. The theory defines four regimes for isoviscous thermal convection, depending on whether kinetic and thermal energy dissipation takes place dominantly in the boundary layer region, or in the convective bulk.

For the ranges of $Ra$ and $Pr$ investigated here, the energy dissipation is dominated by the convective bulk, and the thermal boundary layer is nested inside the kinetic one. In an idealized case such situation yields $Re\propto Ra^{4/9} Pr^{-2/3}$ (see Table 2, regime IV$_u$ in \citet{Ahlers2009}). Based on our 5 data points, we observe $Re = (0.07{\pm}0.01)\, (Ra^{4/9} Pr^{-2/3})^{(1.31{\pm}0.02)}$, i.e.~our exponents differ by ca.~30\% from the idealized case. However, the Grossmann-Lohse theory is derived for a 3D box with thermally insulating side-walls, while we perform 2D simulations with periodic sides, which may be a source of the discrepancy (see also Discussion). When extrapolating to $Ra=10^{27}$ according to our $Re(Ra,Pr)$ relationship, Eq.~\eqref{eqvtvm} gives $\vm\approx[5.2,2.9]$ for $Pr=[10,50]$.

\begin{figure}[ht]
\includegraphics[scale=0.43]{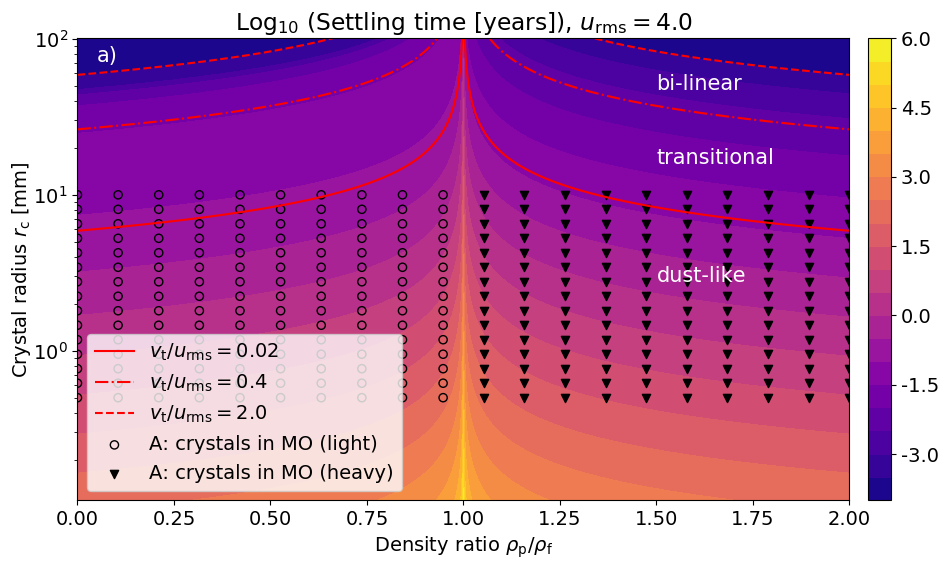}
\includegraphics[scale=0.43]{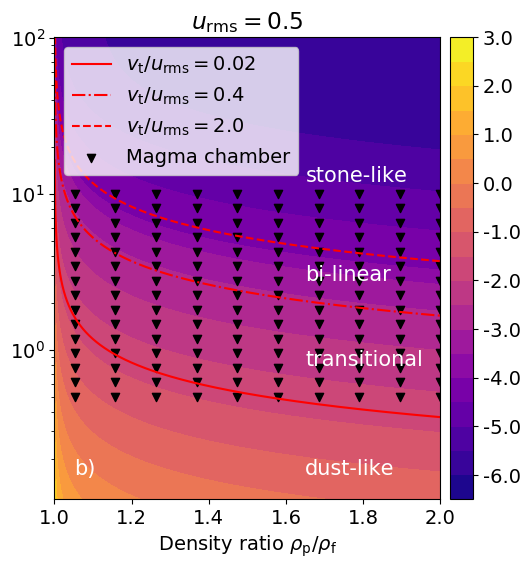}
\caption{(a) Settling time of crystals with $\rc\in [0.1,100]$ mm and $\rho_\mathrm{p}/\rho_\mathrm{f} \in [0,2]$, suspended in a global magma ocean with parameters given in Table \ref{magpar}, for which we estimate $\vm$ to be $[2.9,5.2]$. The black triangles correspond to the black triangles in Fig.~\ref{figFinal}, circles are the respective light particle types. We show $\log_{10}$ of the value of $t_\mathrm{t} F$, where $t_\mathrm{t}$ is the terminal time $H/(\vt u^*)$ and $F$ accounts for the characteristic settling behaviour of the various particle types (see Eq.~\eqref{eqfactor} and Table \ref{tableai}, the parameters $\mu,\sigma,$ and $\lambda$  correspond to the set C$^{12}_{50}$, i.e.~to the simulation with the highest $Ra$). (b) Same as above, only this time $\vm=0.5$ instead of 4.0 and the size of the system is $H=2.9$ km instead of $H=2890$ km. The particle set now spreads across the slow belt. We show only the heavy particles from the set (the position of particle types and the red lines are symmetrical with respect to 1 on the x-axis).}
\label{figMO}
\end{figure}

In Fig.~\ref{figMO}a we show the settling time of crystals in a global magma ocean (Table \ref{magpar}). The results are obtained by multiplying the terminal time $t_\mathrm{t} := 1 / \avt$ (resp.~$H/(\vt u^*)$ in dimensional units) with the factor $F=F(\avt/\vm)$. The black triangles in Fig.~\ref{figMO}a correspond to heavy particles in the particle set A, i.e.~to the black triangles from Fig.~\ref{figFinal}. The mean velocity of the flow is assumed to be $\vm =4.0$, a value representative of the relevant range $[2.9,5.2]$. Since $\vm$ is now larger than in Fig.~\ref{figFinal}, the slow belt moves to the right with respect to the positions of the particle types, which now fall into the dust-like and transitional ranges (see the red lines and black symbols in Figs \ref{figFinal} and \ref{figMO}a).

The factor $F$ is computed using the $a_\mathrm{i}$ coefficients derived for the particle set C$^{12}_{50}$, as this simulation has the highest $Ra$ and includes the $\beta$-effect. $F$ reaches a maximum of 19 for light particles, while for the heavy ones it only slightly exceeds 3 (see the light/heavy asymmetry in Fig.~\ref{figMO}a). 

Note that there are three independent parameters in the simplified Maxey-Riley equation \eqref{eqnonDMR}, but in the dimensional version \eqref{eqMR} there are only two, $\tau_D$ and $\beta$, because the gravitational acceleration $g$ is usually fixed. Considering $St\,\Lambda,$ and $\beta$ as independent is thus a generalization for arbitrary gravity. In Fig.~\ref{figMO} gravity is fixed again, $g=9.8$ m/s$^2$, and the two independent variables are chosen as $\rho_\mathrm{p} / \rho_\mathrm{f}$ and $\rc$, because these quantities are typically measured.

The settling times that we obtained tend to be small compared to the typical lifetimes of magma oceans \citep[e.g.][]{Lebrun2013,Nikolaou2019}. This suggests that a magma ocean will likely solidify via fractional crystallization. An exception is for particle types with extremely small density contrasts. For example, for $\rc=1$ mm, the value of $|\rho_\mathrm{p} / \rho_\mathrm{f} - 1|$ must be smaller than $4\times 10^{-6}$ in order to obtain settling time longer than 1 Myr, while the typical values are ca.~[0.01,0.25] in a cooling magma \citep[e.g.][]{Koyaguchi1990}.

While the position of the slow belt in the ($St, \Lambda$)-diagram depends on the value of $\vm$ only, the positions of crystals of given sizes and radii in that diagram depend on various other parameters (see Eq.~\eqref{eqStLBeta}). Similarly as for a global magma ocean, we can provide first-order estimates also for magma chambers. The key difference between the two is in the value of $H$. When $H=3$ km instead of $H \approx 3\times10^3$ km, crystals with the same radius range $\rc^\mathrm{ref}=[0.5,10]$ mm move by 3 orders of magnitude to the right in Fig.~\ref{figFinal}. At the same time, the lower value of $H$ reduces the Rayleigh number to $\approx\,10^{18}$, yielding $\vm\approx 0.5$, which shifts the slow belt by a factor of 5 to the right in Fig.~\ref{figFinal}. As a result, the crystals span over all the settling regimes, with 0.5 mm crystals being in the dust-like regime for $\rho_\mathrm{p}/\rho_\mathrm{f}$ up to 1.5, and 10 mm crystals being in the stone-like regime for $\rho_\mathrm{p}/\rho_\mathrm{f}$ larger than 1.1 (Fig.~\ref{figMO}b). Again, the expected density contrasts result in settling times significantly smaller than the life-span of the system \citep[typically more than 100 kyr, see e.g.~the review][]{Marsh1989}, indicating fractional crystallization.

The position of the studied particles with respect to the slow belt has important consequences, as it is directly related to the horizontal distribution upon sedimentation. Thus, for a mantle-deep magma ocean, the crystal radius must be $\gtrsim 10$ mm in order to experience a horizontally non-uniform accumulation of sediments (see the black triangles that fall into the transitional regime in Fig.~\ref{figMO}a). For a 3-km-deep magma chamber, on the other hand, the overlap between $[0.5,10]$ mm crystals and the slow belt indicates that the majority of suspended particles will eventually settle in the low-velocity piles (under the assumption that a large scale circulation is present and the low-velocity piles form, see also Discussion).

Note that the rhythmic sedimentation suggested by \citet{Sparks1993} intrinsically relies upon the assumption that precipitated crystals settle much faster in a non-convecting fluid than when convection is present. The existence of the slow-belt presented here is thus in favour of the scenario proposed by \citet{Sparks1993}, for which the authors found petrological evidence in fully solidified chambers. 

\section{Discussion}\label{sec:Discuss}

The crystallization of a primordial molten mantle is a complex system in which the generation, settling, and re-entrainment of crystals are competing processes. Here we only focused on one of these components: the settling of crystals.

Typical time scales for the solidification of a whole-mantle terrestrial magma ocean range from $\sim 10^3$ years in the absence of atmosphere, up to $\sim 10^6$ years in the presence thereof \citep[e.g.][]{Lebrun2013,Nikolaou2019}. When compared to these time scales, our results indicate a fast settling (Fig.~\ref{figMO}a), and thus support the idea of a fully fractional crystallization. In the series of papers by Solomatov \citep[for a review, see][]{Solomatov2015}, it is instead argued for equilibrium crystallization of the majority of the primordial mantle. This is largely because re-entrainment of sedimented particles from the bottom of the fluid is claimed to be the dominant process. \citet{Solomatov1993a} derive a formula for the equilibrium crystal fraction, $\Phi_\mathrm{eq} = 18 \epsilon \alpha \rho \nu Q / (g c_\mathrm{p} \Delta \rho^2 r_\mathrm{c}^2)$, where $Q$ is the surface heat flux and $\epsilon$ denotes the fraction of available convective energy that goes into re-entrainment, estimated to be $\approx 0.1 - 1$ \% \citep{Solomatov1993a}. For the simulations presented here, e.g.~for $Ra=10^{10},\,Pr={50}$ and crystals with $\Delta \rho/\rho=0.1$ and $r_\mathrm{c}=1$ cm, the resulting $\Phi_\mathrm{eq}$ is only around 3 \% (the formula for $\Phi_\mathrm{eq}$ is designed for a single type of crystals only; for a range of crystal properties it must be decided how much of the available energy goes into the lifting of the various types). For the significantly larger heat fluxes that accompany the early stages of a global magma ocean solidification, it quickly reaches 100\%, indicating full suspension \citep{Solomatov1993c}.

More recently, the scaling law of \citet{Solomatov1993a} was confirmed by the experimental study of \citet{Lavorel2009}, who systematically varied the density ratio $\rho_\mathrm{p}/\rho_\mathrm{f}$ and the temperature contrast that drives thermal convection. An important finding of their study is that in a highly turbulent flow the molecular viscosity that appears in the formula for the Stokes' velocity must be replaced by an apparent viscosity in order to account for turbulent eddies smaller than the particles. Such approach was successfully used to describe the dynamics of finite-size particles in turbulent flows \citep{Qureshi2007,Brito2004}, but it is difficult to generalize in non-homogeneous flows such as for thermal convection between parallel plates. In the context of particle settling, \citet{Lavorel2009} found this approach viable, and the apparent viscosity that they measured at $Ra=3\times 10^9$ was only ca.~2.7 times larger than the molecular viscosity. We note, however, that replacing molecular viscosity with apparent (or turbulent) viscosity in their formula for the decay of the number of suspended particles (their Eq.~(9)) is similar to dividing the terminal velocity $\avt$ by $f$ as in our Eq.~\eqref{eqOurmodel}, resp.~it is mathematically identical if $f$ can be treated as constant for the investigated particles. Moreover, the amplitude 2.7 is within the range that we obtain for $f$. In other words, the decrease of the Stokes' velocity that \citet{Lavorel2009} computed to reconcile their measurements may have been caused by the effects of large-scale circulation analyzed in this paper as well as by the effect of sub-particle sized turbulence. 

Similarly to the previous works, \citet{Lavorel2009} used a cuboid tank filled with salty water and spherical, polymethyl methylacrylate particles. We note that the conditions of the experiments \citep{Olson1984, Martin1989, Solomatov1993a} on which the energetic analysis of \citet{Solomatov1993b} is based differ significantly from the environment of non-spherical silicate crystals that accumulate at the bottom of a cooling magma. In nature, the sedimented crystals may be subject to chemical and petrological altering, possibly binding the crystals together, i.e.~making them prone to re-entrainment. Note that fractional crystallization is often reported in exposed plutons \citep[e.g.][]{Sparks1993}. While we do not argue against re-entrainment as such, we merely point out that its workings should be thoroughly investigated in future work in the context of magma environments.

In fact, the non-uniform horizontal distribution of settling events that we observe in the transitional and bi-linear regimes is slightly in favour of re-entrainment. As discussed in \citet{Solomatov1993a}, embedded particles may be lifted by the tangential stresses caused by rising plumes, and the respective stresses increase with distance from the domain boundary. We observe a large concentration of sedimented particles in the low-velocity piles. Inside these regions, the sediments would thus build tall dunes. Since most of young plumes are born at the edges of the low-velocity piles, the crests of these dunes should be exposed to large tangential stresses.

In terms of the predicted settling time, some of our results differ dramatically from those of \citet{Verhoeven2009}. Roughly speaking, for particles satisfying $(\rho_\mathrm{p} - \rho_\mathrm{f}) < \rho_\mathrm{f} \alpha \Delta T $ (see their Fig.~12 and Eq.~5), they obtain a temperature-dominated convection mode (T-regime), in which the flow is thermally driven and all particles are held indefinitely in suspension \rev{(similar results are obtained in the non-rotating cases of \citet{Maas2015,Maas2019})}. This is because in their formulation the momentum equation is solved for the volumetric average of the fluid and particle velocity, i.e.~their particles by definition follow the fluid as described in Eq.~6 in \citet{Verhoeven2009} (but they can also invoke fluid motion, see the next paragraph). Therefore, in case of vigorous thermal mixing of the fluid, their particles never settle. In our study, particles can have different velocities from the surrounding fluid, which allows fluid-particle separation and thus sedimentation regardless of convective vigor.

For stronger density contrasts, roughly for $(\rho_\mathrm{p} - \rho_\mathrm{f}) > \rho_\mathrm{f} \alpha \Delta T $, \citet{Verhoeven2009} obtain a particle-driven convection in which a layer of sediment is segregated from the rest of the fluid (C-regime). We note that the condition $(\rho_\mathrm{p} - \rho_\mathrm{f}) > \rho_\mathrm{f} \alpha \Delta T $ simply means that the critical concentration that is required for the formation of a settling front, as described earlier by \citet{Koyaguchi1990} and \citet{Sparks1993}, is less than 100\%. In other words, the C-regime is established whenever the formation of a settling front (and thus cessation of convection due to particle motion) can take place for some critical particle concentration $C_* < 100$ \%. As analyzed by \citet{Solomatov1993b}, for the crystals of interest in magma oceans and chambers, $C_*$ is typically less than 100\% (and this is also the case for most of the particle types investigated in the present paper).  

Within the C-regime, \citet{Verhoeven2009} develop a model that is based on the theory of \citet{Martin1989}. In particular, they complement the theory by accounting for the volume occupied by particles that have already sedimented. This is not to be confused with our model, in which the function $f$ is a measure of the rate of particle transport into the low-velocity piles, normalized by the probability of not escaping from these regions. As such, our function $f$ depends non-trivially on the structure of the background flow, while the factor $f$ in Eqs.~26--32 of \citet{Verhoeven2009} represents packing of sedimented material. \citet{Verhoeven2009} then verify their model on a set of simulations in which the employed particles fall into the dust-like regime according to our classification (see their Eq.~8, Fig.~10, and the parameters listed below Fig.~15). This should be understood as yet another confirmation of the applicability of Eq.~\eqref{eqNokes} for particles with a small value of the $\avt/\vm$ ratio. 

At the beginning of our simulations, we inject the particles uniformly throughout the entire model domain. In systems where the particulate phase is a product of a chemical reaction or phase change, this is typically not the case. In particular, in a cooling magma, the solid crystals nucleate in the relatively cold downwellings and degassing takes place due to decompression in hot rising plumes. The importance of the initial positions of newly formed particles depends on their $\vt/\vm$ ratio. On one hand, in the dust-like regime the particles are likely to get thoroughly mixed and the settling rates would not be affected. In the stone-like regime, on the other hand, the particles' trajectories and settling times strongly depend on the particles' initial positions and velocities. Given the particle positions associated with the second stage of the bi-linear regime (Fig.~\ref{figBilinear}), faster settling than reported here is to be expected on average if the majority of heavy crystals form preferentially in downwellings.

For $Ra=10^{12}$ and $Pr=50$, i.e.~for the highest investigated Reynolds number, the Kolmogorov length scale $\eta$ is $\approx 2{\times}10^{-3}$, while the upper bound of the non-dimensional particle radius $\rc/H$ is $1.5{\times}10^{-3}$ (i.e.~the smallest turbulent eddies are only slightly larger than the investigated particle radii in the respective simulation set). For the simulation set xC$^{10}_{50}$ the assumed crystal sizes even exceed the Kolmogorov length, and the particle Reynolds numbers are in turbulent rather than laminar regime (i.e.~breaching the range of validity of Eq.~\eqref{eqnonDMR}). We note, however, that the set xC$^{10}_{50}$ is performed only for illustrative purposes in the context of explaining some aspects of the stone-like regime.  

Our study shows that the existence of a stable large-scale flow structure has a clear signature on the settling of particles. First, it delays the settling on average. Second, it is responsible for the non-uniform horizontal distribution of settling events. Since the large-scale circulation is observed at the highest Rayleigh numbers reached so far in numerical simulations \citep{Zhu2018} and experiments \citep{Ahlers2009}, we speculate that this can be an important feature for the extreme regime of a cooling magma ocean. Superstructures were analyzed in detail in 3D systems with large aspect ratios by \citet{Pandey2018}. To test the influence of a larger aspect ratio, we performed the reference simulation C$^{10}_{50}$ also with aspect ratio 4, and the resulting function $f$ as well as the amplitude of the horizontal variations of settling events were nearly identical.

In Section \ref{sec:MO} we performed an extrapolation to $Ra=10^{27}$, assuming $Re = 0.07\, Ra^{0.582} Pr^{-0.873}$. \citet{Grossmann2000} distinguish four idealized regimes of convection depending on whether kinetic and thermal dissipation rates are dominated by the convective bulk or the boundary layers. For our Rayleigh and Prandtl numbers, the idealized case (``pure power-law'') yields $Re\propto Ra^{4/9} Pr^{-2/3}$, but already for $Pr\approx 1$ we would be on the boundary with the $Re\propto Ra^{1/2} Pr^{-1/2}$ regime, in which the kinetic energy dissipates in the velocity boundary layer of the flow. While these are idealized cases, real convection is a mixture of these regimes, and for a detailed treatment one must employ the full theory of \citet{Grossmann2000}. Using scaling laws to extrapolate to very high Rayleigh numbers, such as done in Section \ref{sec:MO}, is, however, still subject to an open debate \citep[e.g.][]{Ahlers2009}.

It is interesting to note that the amplitude of $\vm$ does not depend on $Ra$ and $Pr$ in the limit of $Re\propto Ra^{1/2} Pr^{-1/2}$, i.e.~in the idealized scenario for low Prandtl and high Rayleigh numbers. In such a case, the settling behaviour could be predicted simply by computing the terminal velocities of particles of interest since the $\vt/\vm$ ratio would not depend on the exact values of $Ra$ and $Pr$ (see Eq.~\eqref{eqvtvm}). However, for low values of $Pr$ new settling regimes may exist, which we plan to investigate in the future.

Convection in magma oceans occurs in the presence of rotation, which is not included in our model. \rev{Using estimated values for Earth's magma ocean, the convective Rossby number lies in a range of 0.03 - 100 \citep{Maas2015}}, and the large-scale circulation thus may be disrupted due to rotation \citep{Kunnen2008, Stevens2012}. Both the settling rates and the horizontal distribution of sedimented material would be affected in such scenario. Recently, the effects of rotation on the distribution of crystals and the rate of settling in the early stages of Earth's primordial magma ocean were analyzed in spherical geometry by \citet{Maas2019}. \rev{In a rotation dominated scenario (Rossby $\lesssim 0.3$) they find a much pronounced settling as convection and thus vertical entrainment of particles is suppressed.}

\section{Summary}\label{sec:Summary}
We evaluate the settling rate of inertial particles that are injected into statistically steady state thermal convection. Previously, the number of suspended particles in such system was assumed to follow either the relation $N=N_0\exp(-\vt t)$ or $N=N_0(1-\vt)$, with $\vt$ being the Stokes' velocity. We observe a new regime with particularly slow sedimentation, in which large-scale circulation prevents particles from reaching the boundary layers of the fluid. By introducing a new framework that treats the settling mechanism as a random process, we develop a model that unifies the observed settling rates into the general equation $N=N_0\exp(-\vt\,t / f)$, where $f$ is a function of the ratio of Stokes' and mean characteristic velocity of the flow, $\vt/\vm$. We investigate $f$ over a broad range of Reynolds numbers and show that the function is relatively robust. It reaches its maximum for $\vt/\vm \approx 0.4$, the maximum value ranging approximately from 1.5 to 3 for particles with mild density contrasts with respect to density of the fluid (Table \ref{tableai}a).

We also analyze the horizontal distribution of settled particles. Within the regime of slow settling, heavy particles accumulate preferentially below major clusters of upwellings. These are located at edges of large-scale convection rolls.

For Reynolds numbers larger than $\sim \rev{1000}$ and particles with a stronger density contrast, additional complexity arises because of the preferential concentration phenomenon, i.e.~light particles have a tendency to move towards flow vortices, while heavy particles move away from them. As a result, light particles get captured inside long-lived vortices, which significantly prolongs their sedimentation at the top boundary. The maximum value of $f$ is close to 10 (resp.~the normalized settling time $F \approx 30$) for our simulation with the highest Rayleigh number, $Ra=10^{12}$.

When extrapolated to the extreme conditions of solidifying magma chambers and oceans, our results predict fractional crystallization. For a better understanding of such complex systems, it is possible (and necessary) to extend our method to account for 3D geometry, rotation, re-entrainment of sedimented particles, self-consistent nucleation of solid crystals, and the coupling between particle concentration and momentum conservation of the fluid.

\section{Acknowledgement}
\rev{We thank Tina R\"{u}ckriemen for useful discussions. VP and NT acknowledge support from the Helmholtz Association (project VH-NG-1017) and the DLR-DAAD research fellowship program.} 

\noindent

\bibliographystyle{model1-num-names}
\bibliography{bibfile.bib}


\end{document}